\documentclass[amsmath,amssymb,showpacs]{revtex4}
\usepackage{graphicx}
\usepackage{dcolumn}
\usepackage{bm}
\makeatletter \@addtoreset{equation}{section} \makeatother

\newcommand{\e}{\epsilon}

\begin{document}


\title{Two-dimensional defect modes in optically induced photonic lattices}
\author{Jiandong Wang, Jianke Yang}
\email{jyang@math.uvm.edu} \affiliation{Department of Mathematics
and Statistics, University of Vermont, Burlington, VT 05401, USA}

\author{Zhigang Chen}
\affiliation{Department of Physics and Astronomy, San Francisco
State University, San Francisco, CA 94132, USA
}%

\begin{abstract}
In this article, localized linear defect modes due to bandgap
guidance in two-dimensional photonic lattices with localized or
non-localized defects are investigated theoretically. First, when
the defect is localized and weak, eigenvalues of defect modes
bifurcated from edges of Bloch bands are derived analytically. It is
shown that in an attractive (repulsive) defect, defect modes
bifurcate out from Bloch-band edges with normal (anomalous)
diffraction coefficients. Furthermore, distances between defect-mode
eigenvalues and Bloch-band edges are exponentially small functions
of the defect strength, which is very different from the
one-dimensional case where such distances are quadratically small
with the defect strength. It is also found that some defect-mode
branches bifurcate not from Bloch-band edges, but from quasi-edge
points within Bloch bands, which is very unusual. Second, when the
defect is localized but strong, defect modes are determined
numerically. It is shown that both the repulsive and attractive
defects can support various types of defect modes such as
fundamental, dipole, quadrupole, and vortex modes. These modes
reside in various bandgaps of the photonic lattice. As the defect
strength increases, defect modes move from lower bandgaps to higher
ones when the defect is repulsive, but remain within each bandgap
when the defect is attractive, similar to the one-dimensional case.
The same phenomena are observed when the defect is held fixed while
the applied dc field (which controls the lattice potential)
increases. Lastly, if the defect is non-localized (i.e. it persists
at large distances in the lattice), it is shown that defect modes
can be embedded inside the continuous spectrum, and they can
bifurcate out from edges of the continuous spectrum algebraically
rather than exponentially.

\end{abstract}


\maketitle

\section{Introduction}
Study of light propagation in periodic media such as photonic
crystals and optically induced photonic lattices is of great
interest to both fundamental physics and applications \cite{Russell,
Joannopoulos,Christodoulides_review,Kivshar_book}. It is well known
that the unique feature of such periodic systems is the existence of
bandgap structures in its linear spectrum. Inside the bands,
eigenmodes of its spectrum are Bloch waves, while inside bandgaps,
wave propagation is forbidden because of the repeated Bragg
reflections. To guide light in periodic media, one of the convenient
ways is to introduce a defect into the medium. The defect can
support linear localized modes (called defect modes) inside bandgaps
of the periodic medium. Such modes could only propagate along the
defect direction (their propagation along other directions is
forbidden for being inside bandgaps), thus defect guidance (or
bandgap guidance) of light is realized.

Defects and the corresponding defect modes (DMs) have been widely
investigated in the field of photonic crystals \cite{Russell,
Joannopoulos}, where bandgaps and defect modes are in the temporal
frequency domains. Recently, reconfigurable optically induced
photonic lattices in photorefractive crystals with and without
defects were successfully generated
\cite{Segev_nature,ChenPRL04,ChenYangPRL,Chen_OE}. In photonic
lattices, bandgaps are in the spatial frequency domains. Linear DMs
in one-dimensional (1D) photonic lattices have been analyzed both
theoretically and experimentally in
\cite{YangOL,YangChendefect,Chen_OE}. Nonlinear defect solitons in
1D photonic lattices have also been theoretically explored
\cite{Kivshar_defect,YangChenPRE} (see also a study of DMs and
defect solitons in 1D waveguide arrays in \cite{Silberberg}). A more
interesting subject is DMs in 2D photonic lattices, where richer
light-guiding possibilities can arise. Recently, we have observed
that a localized defect in 2D lattices can guide not only
fundamental modes, but also higher-order modes with delicate tail
structures \cite{ChenYangPRL}. However, our theoretical
understanding on such new types of 2D DM structures is still far
from satisfactory. For instance, it is still unclear what types of
DMs a 2D defect can possibly support, and how localized these DMs
can be in defects of various depths. We should mention that in
uniformly periodic photonic lattices, a number of nonlinear
localized modes such as fundamental, dipole, vortex, dipole-array
and vortex-array solitons have been reported
\cite{Eisenberg98,Segev_nature,ChenPRL04,
Neshev_dipole,Yang_dipole,Yang_Muss,Chen_vortex, Segev_vortex,
Segev_highervortex,Kivshar_C_onemode,Shi}. Solitons in Bessel-ring
lattices and 2D quasi-periodic lattices have been reported as well
\cite{Kartashov,Chen_ring,Ablowitz}. These nonlinear modes may be
regarded as linear DMs of the defect induced by solitons themselves.
Hence studies of linear DMs in defective lattices and nonlinear
solitons in uniform lattices can stimulate each other. Indeed, many
types of linear DMs we will report in this paper do resemble certain
types of strongly localized solitons in 2D uniform lattices (see
\cite{Segev_highervortex,Kivshar_C_onemode,Shi}). However, many
differences exist between solitons in uniform lattices and linear
DMs in defective lattices. For instance, low-amplitude solitons in
uniform lattices are very broad and occupy many lattice sites, thus
the ``defects" these solitons generate are very different from the
single-site defects as created in the experiments of
\cite{ChenYangPRL} and considered in several sections of this paper.
Another difference is that, in uniform lattices, centers of solitons
can be located at two positions (on-site and off-site) in the 1D
case \cite{Peli_1D} and four positions at or between lattice sites
in the 2D case \cite{Shi}. However, in defective lattices, linear
DMs can only be centered at the defect site. Due to these
differences, previous results on solitons in uniform lattices can
not be naively copied over to linear DMs.

In this article, linear localized DMs in 2D optically induced
photonic lattices with localized or non-localized defects are
systematically studied theoretically. For localized defects, we show
numerically that both the attractive and repulsive defects can
support various types of 2D DMs such as fundamental, dipole,
quadrupole, and vortex modes. These modes reside in various bandgaps
of the photonic lattice. When the defect is weak, DMs are further
studied analytically by asymptotic methods. We show that in a weak
repulsive defect, DMs bifurcate out from Bloch-band edges with
anomalous diffraction coefficients. The situation is opposite for
attractive defects. We also show that distances between DM
eigenvalues and Bloch-band edges are exponentially small functions
of the defect strength, which is very different from the 1D case
where such dependence is quadratic. In addition, we find that some
DM branches bifurcate not from Bloch-band edges, but from quasi-edge
points within Bloch bands, which is very unusual. As the defect
strength increases, DMs move from lower bandgaps to higher ones when
the defect is repulsive, but remain within each bandgap when the
defect is attractive. The same phenomena are observed when the
defect is held fixed while the applied dc field (which controls the
lattice potential) increases. If the defect is non-localized (i.e.
it does not disappear at large distances in the lattice), we show
that DMs exhibit some new features, such as DMs can be embedded
inside the continuous spectrum, and they can bifurcate out from
edges of the continuous spectrum algebraically rather than
exponentially.

\section{The Theoretical Model}
We begin our study by considering one ordinarily polarized 2D
square-lattice beam with a \emph{localized} defect and one
extraordinarily polarized probe beam with very low intensity
launched simultaneously into a photorefractive crystal. The two
beams are mutually incoherent, and the defective lattice beam is
assumed to be uniform along the propagation direction (such
defective 2D lattices have been created in our earlier experiments
\cite{ChenYangPRL}). In this situation, the dimensionless governing
equation for the probe beam is \cite{Segev_nature,
Christodoulides_model, Segev_model, YangChendefect}:
\begin{equation}
 iU_z+ U_{xx}+U_{yy}-\frac{E_0}{1+I_L(x,y)}U=0. \label{eq:one}
\end{equation}
Here $U$ is the slowly-varying amplitude of the probe beam, $(x, y)$
are transverse distances (in units of $D/\pi$ with $D$ being the
lattice spacing), $z$ is the propagation distance (in units of
$2k_1D^2/\pi^2$), $E_0$ is the applied dc field (in units of
$\pi^2/(k_0^2n_e^4D^2r_{33})$),
\begin{equation}
I_L=I_0\cos^2(x)\cos^2(y)[1+\epsilon F_D(x,y)] \label{eq:two}
\end{equation}
is the intensity function of the photonic lattice (normalized by
$I_d+I_b$, where $I_d$ is the dark irradiance of the crystal and
$I_b$ the background illumination), $I_0$ is the peak intensity of
the otherwise uniform photonic lattice, $F_D(x,y)$ describes the
shape of the defect, $\epsilon$ controls the strength of the defect,
$k_0=2\pi/\lambda_0$ is the wave number ($\lambda_0$ is the
wavelength), $k_1=k_0 n_e$, $n_e$ is the unperturbed refractive
index, and $r_{33}$ is the electro-optic coefficient of the crystal.
Notice that the period of the square lattice has been normalized to
be $\pi$. For localized defects which we are considering in this
section as well as sections \ref{DMbifurcation} and
\ref{DC_dependence}, $F_D(x,y)$ is a localized function (the case of
non-localized defects will be considered in Sec.
\ref{non-localized}). In our numerical computations, for simplicity,
we choose $F_D(x,y)$ to be \cite{YangChendefect,ChenYangPRL}
\begin{equation} \label{FD}
F_D(x,y)=\exp[-(x^2+y^2)^4/128],
\end{equation}
which describes a single-site defect. When $\epsilon>0$, the lattice
intensity $I_L$ at the defect site is higher than that at the
surrounding regions, and a defect like this is called an attractive
defect. Otherwise, the defect is called a repulsive one. The
defected lattice profiles with $\epsilon=\pm 0.9$ will be displayed
later in the text (see Figs. \ref{figure5} and \ref{figure7}).
Throughout this paper (except in section \ref{non-localized}), we
will choose the lattice peak intensity to be $I_0=6$.

Localized defect modes in Eq.~(\ref{eq:one}) are sought in the form
of
\begin{equation}
U(x,y;z)=u(x,y)\exp(-i\mu z), \label{eq:three}
\end{equation}
where $\mu$ is the propagation constant, and $u(x,y)$ is a localized
function in $x$ and $y$. After substituting the above expression
into Eq.~(\ref{eq:one}), a linear eigenvalue equation for $u(x,y)$
is derived:
\begin{equation}
u_{xx}+u_{yy}+[\mu-\frac{E_0}{1+I_L(x,y)}]u=0. \label{eq:four}
\end{equation}
In the rest of this paper, we will comprehensively analyze DMs in
this 2D Schr\"odinger equation with defective lattice potentials by
both analytical and numerical techniques.

\begin{figure}
\centering
\begin{minipage}[c]{0.6\textwidth}
\centering
\includegraphics[width=0.92\textwidth]{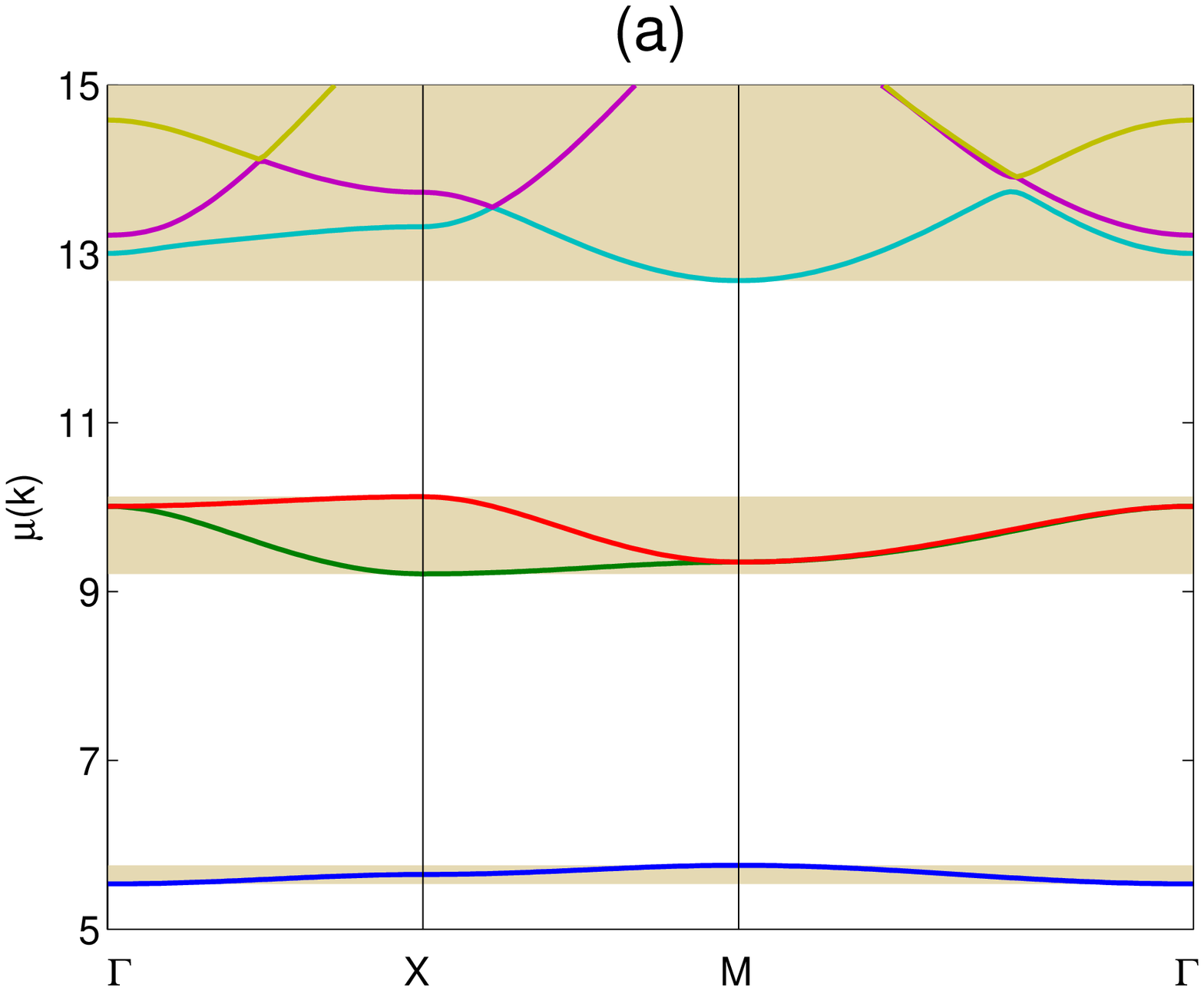}
\end{minipage}
\begin{minipage}[c]{0.2\textwidth}
\centering
\includegraphics[width=0.9\textwidth]{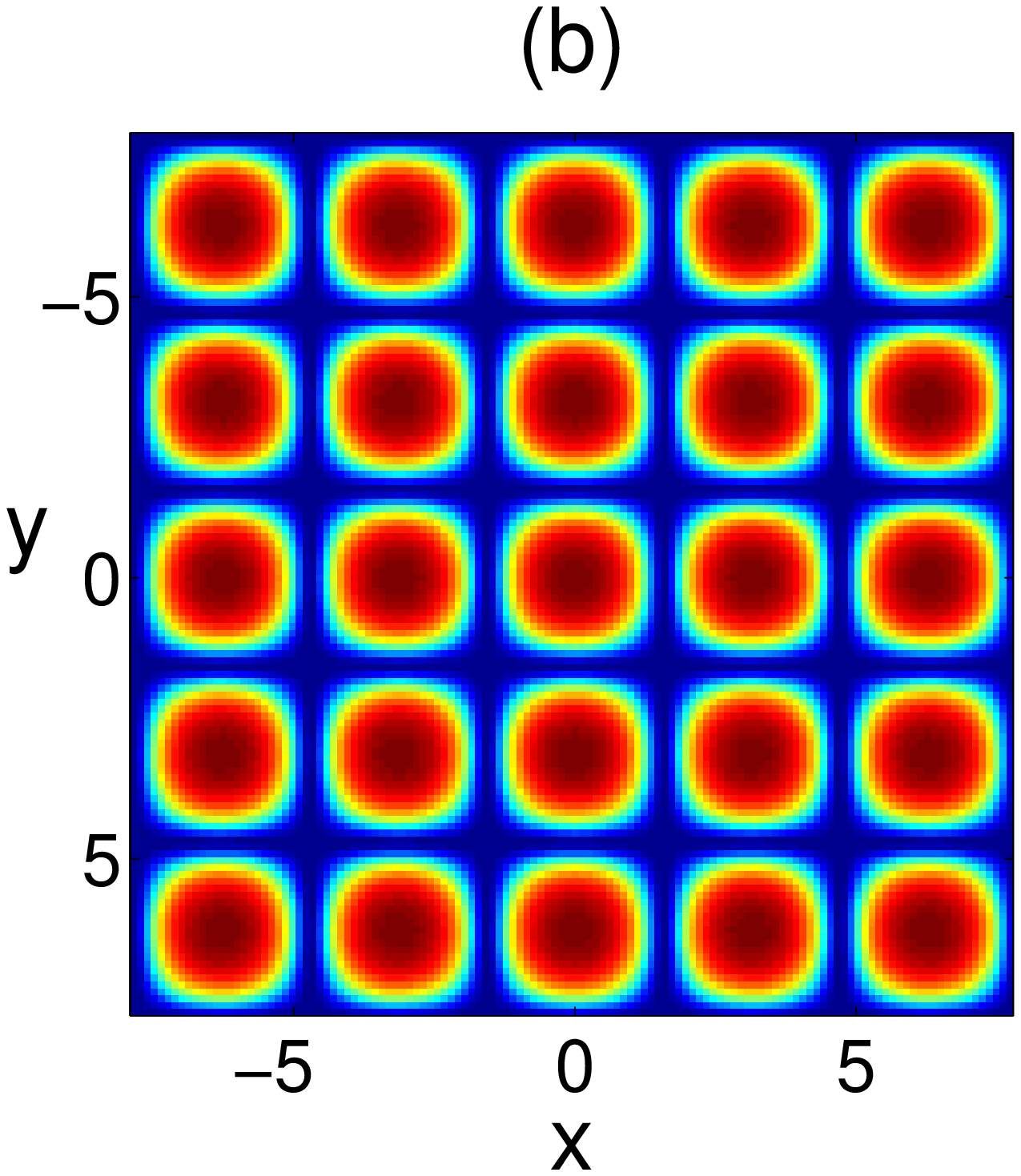}

\vspace{0.5cm}
\includegraphics[width=0.95\textwidth]{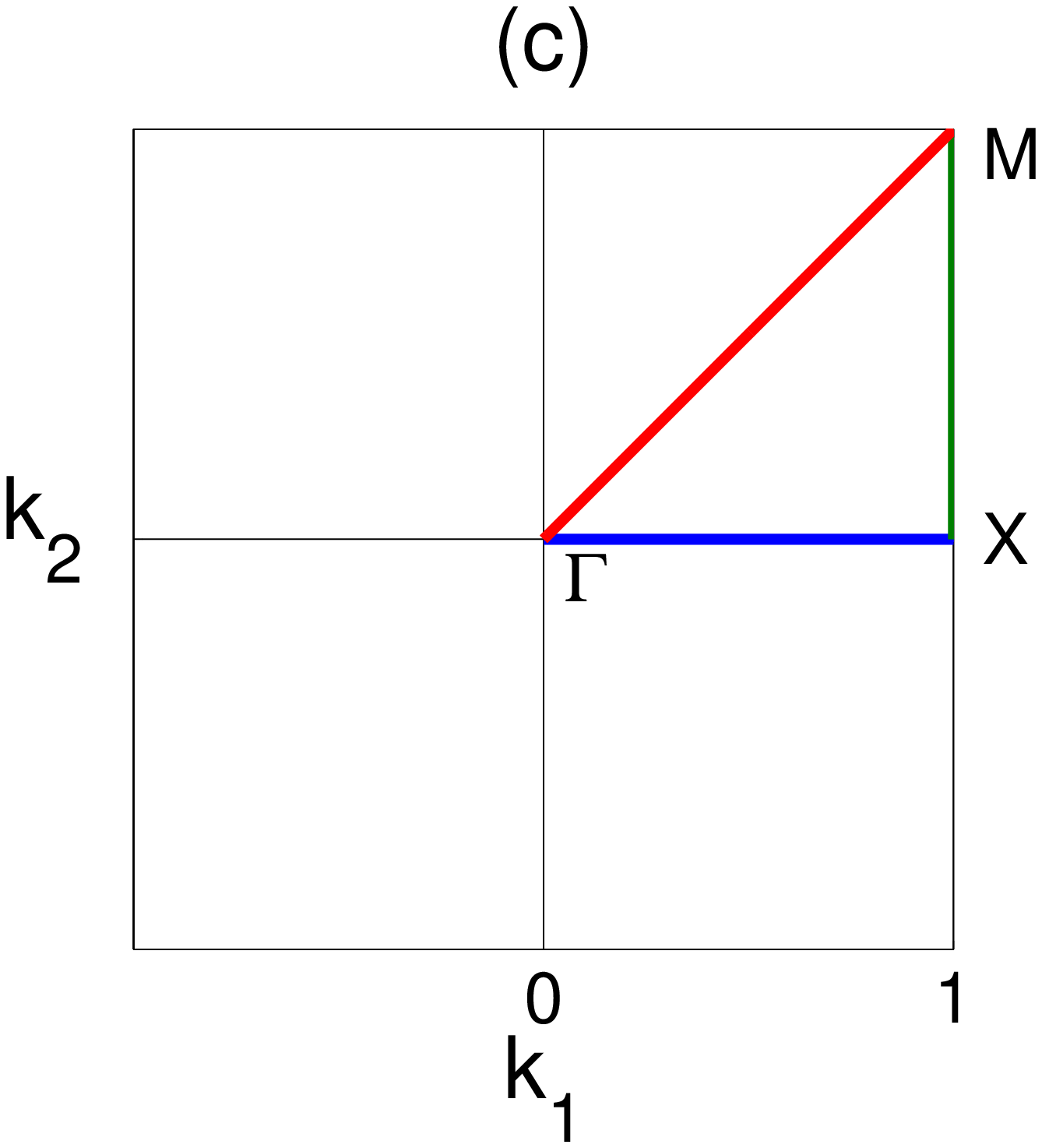}
\end{minipage}
\caption{(a) Diffraction relation of the uniform lattice potential
(\ref{V}) with $E_0=15$ and $I_0=6$ in the reduced Brillouin zone
along the direction of $\Gamma\rightarrow X\rightarrow
M\rightarrow\Gamma$. Shaded: first three Bloch bands. (b) Potential
(i.e. effective refractive index change) (\ref{V}) in the $(x, y)$
plane. (c) The first Brillouin zone of the 2D lattice in the
reciprocal lattice space.} \label{figure1}
\end{figure}

Before the analysis of DMs, let us first look at the diffraction
relation and bandgap structure of Eq.~(\ref{eq:four}) without
defects (i.e., $\epsilon=0$). According to the Bloch theorem,
eigenfunctions of Eq.~(\ref{eq:four}) can be sought in the form of
\begin{equation}
u(x, y)= e^{ik_1x+ik_2 y} G(x, y; k_1, k_2), \quad \mu=\mu(k_1,
k_2), \label{eq:five}
\end{equation}
where $\mu=\mu(k_1, k_2)$ is the diffraction relation, wavenumbers
$k_1$, $k_2$ are in the first Brillouin zone, i.e. $-1 \le k_1, k_2
\le 1$, and $G(x, y; k_1, k_2)$ is a periodic function in $x$ and
$y$ with the same period $\pi$ (in normalized units) as the uniform
lattice of (\ref{eq:two}). By substituting the above Bloch solution
into Eq. (\ref{eq:four}) (with $\e=0$), we find that the diffraction
relation $\mu(k_1, k_2)$ of our 2D uniform lattice will be obtained
by solving the following eigenvalue problem
\begin{equation}
[(\partial_x+ik_1)^2+(\partial_y+ik_2)^2+V(x,y)]G(x, y)=-\mu G(x, y)
\label{eq:six}
\end{equation}
with the uniform periodic potential
\begin{equation} \label{V}
V(x,y)=-\frac{E_0}{1+I_0 \cos^2(x)\cos^2(y)}.
\end{equation}
Figure \ref{figure1} shows the diffraction relation of
Eq.~(\ref{eq:four}) along three characteristic high-symmetry
directions ($\Gamma\rightarrow X\rightarrow M\rightarrow\Gamma$) of
the irreducible Brillouin zone. It can be seen that there exist
three complete gaps which are named the semi-infinite, first and
second gaps respectively. They correspond to the white areas in
Figure \ref{figure1} from the bottom to the top, separated by the
shaded Bloch bands.

Loosely speaking, a lattice with an attractive defect is like a
uniform lattice superimposed with a bright beam (soliton) under
self-focusing nonlinearity, while a lattice with a repulsive defect
corresponds to that under self-defocusing nonlinearity. It is well
known that spatial solitons can be formed if the diffraction of the
medium is normal (corresponding to $\mu''(k)>0$ in our current
notations), and the nonlinearity of the medium is self-focusing; or
the diffraction is anomalous ($\mu''(k)<0$) and the nonlinearity is
self-defocusing. From Figure \ref{figure1}(a), we can see that the
diffraction relation $\mu(k_1, k_2)$ is normal at the lower edge of
every band, and anomalous at the upper edge of every band. Thus it
can be anticipated that in a weak attractive defect, DMs will
bifurcate out from the lower edge of every Bloch band; while in a
weak repulsive defect, DMs will bifurcate out from the upper edge of
every Bloch band. For 1D defects, this heuristic argument has been
fully confirmed in \cite{YangChendefect}. For 2D defects, it will be
validated in this paper as well. Of course, such heuristic arguments
can not offer us any insight on quantitative behaviors of DM
bifurcations. In the 1D case, it has been shown that eigenvalues of
DMs depend on the defect strength $\epsilon$ quadratically when the
defect is weak \cite{YangChendefect}. For 2D defects, DM eigenvalues
turn out to be exponentially small with the defect strength
$\epsilon$ (see below), which is distinctively different from the 1D
case. In the next section, we will study in detail how DMs bifurcate
from edges of Bloch bands both analytically and numerically. Another
interesting phenomenon which we will demonstrate there is that some
DM branches do not bifurcate from Bloch-band edges. Rather, they
bifurcate from points \emph{inside} Bloch bands.

\section{Dependence of defect modes on the strength $\e$ of localized defects} \label{DMbifurcation}

In this section, we consider DMs in the model equations
(\ref{eq:one})-(\ref{eq:two}) under \emph{localized} defects, and
investigate how they depend on the defect strength $\e$. First, we
analytically study how DMs bifurcate from edges of Bloch bands when
the defect is weak. The method we will use is analogous to one used
in \cite{Peli_exp} for eigenvalue bifurcations in the Schr\"odinger
equation with a weak radially symmetric potential. Second, we
numerically determine various types of DMs under both weak and
strong defects of the form (\ref{FD}) by directly solving the 2D
eigenvalue problem (\ref{eq:four}). The numerical method we use is a
power-conserving squared-operator iteration method developed in
\cite{YangLakoba}.

\subsection{Bifurcations of defect modes from Bloch-band edges in weak localized defects}

Consider the general two-dimensional perturbed Hill's equation
\begin{equation}
\psi_{xx}+\psi_{yy} + \left[ \mu+V(x, y) \right] \psi =\epsilon f(x,
y)\psi, \label{perturb_eq}
\end{equation}
where the unperturbed potential function $V(x, y)$ is periodic along
both the $x$ and $y$ directions with periods $T_1$ and $T_2$
respectively, the perturbation to the potential, $f(x, y)$, is a
2D-localized defect function in the $(x, y)$ plane (i.e. $f(x, y)
\to 0$ as $(x, y) \to \infty$), and $\epsilon \ll 1$. Application of
our general results on Eq. (\ref{perturb_eq}) to the specific case
of Eq. (\ref{eq:four}) will be given in the end of this subsection.
Note that the 2D-localization assumption on the function $f(x, y)$
is important for the analysis below. If $f(x, y)$ is not
2D-localized, then DM bifurcation behaviors could be quite different
(see section \ref{non-localized}).

When $\epsilon=0$, Eq. (\ref{perturb_eq}) admits Bloch solutions of
the form
\begin{equation}
\psi(x, y)= B_n(x, y; k_1, k_2) \equiv e^{ik_1x+ik_2 y} G_n(x, y;
k_1, k_2), \quad \mu=\mu_n(k_1, k_2),
\end{equation}
where $\mu=\mu_n(k_1, k_2)$ is the diffraction relation of the
$n$-th Bloch surface, ($k_1$, $k_2$) lies in the first Brillouin
zone, i.e. $-\pi/T_1 \le k_1 \le \pi/T_1$, $-\pi/T_2 \le k_2 \le
\pi/T_2$, and $G_n(x, y; k_1, k_2)$ is a periodic function in both
$x$ and $y$ with periods $T_1$ and $T_2$ respectively. All these
Bloch modes $\{B_n(x, y; k_1, k_2), (k_1, k_2) \in \mbox{the
Brillouin zone}, n=1, 2, \dots \}$ form a complete set
\cite{keller}. In addition, the orthogonality condition between
these Bloch modes is
\begin{equation}  \label{orth}
\int_{-\infty}^\infty \int_{-\infty}^\infty B_m^*(x, y; k_1,
k_2)B_n(x, y; \hat{k}_1, \hat{k}_2)dxdy =(2\pi)^2\:
\delta(k_1-\hat{k}_1) \delta(k_2-\hat{k}_2) \delta(m-n).
\end{equation}
Here the Bloch functions have been normalized by
\[\frac{\int_0^{T_1}dx\int_0^{T_2}dy |G_n(x, y; k_1, k_2)|^2}{T_1T_2}=1,\]
$\delta(.)$ is the $\delta$-function, and the superscript ``*''
represents complex conjugation. When $\epsilon \ne 0$, localized
eigenfunctions (i.e. DMs) can bifurcate out from edges of Bloch
bands into band gaps. Asymptotic analysis of these DMs for $\epsilon
\ll 1$ will be presented below.

Consider bifurcations of defect modes from an edge point $\mu=\mu_c$
of the $n$-th Bloch diffraction surface. If two or more Bloch
diffraction surfaces share the same edge point, each diffraction
surface would generate its own defect mode upon bifurcation, thus
one could treat each diffraction surface separately. After
bifurcation, the propagation constant $\mu$ of the defect mode will
enter the band gap adjacent to the diffraction surface. Without loss
of generality, we assume that the edge point of this diffraction
surface is located at a $\Gamma$ symmetry point of the Brillouin
zone where $(k_1, k_2)=(0, 0)$. If the bifurcation point is located
at other points such as $M$ or $X$ points, the analysis would remain
the same.

When $\epsilon \ne 0$, defect modes can be expanded into Bloch waves
as
\begin{equation}
\psi \left( x, y \right) =\sum_{n=1}^\infty
\int_{-\pi/T_1}^{\pi/T_1} dk_1 \int_{-\pi/T_2}^{\pi/T_2} dk_2 \
\alpha_n (k_1, k_2)B_n(x, y; k_1, k_2), \label{expansion}
\end{equation}%
where $\alpha_n (k_1, k_2)$ is the Bloch-mode coefficient in the
expansion. In the remainder of this subsection, unless otherwise
indicated, integrations for $dk_1$ and $dk_2$ are always over the
first Brillouin zone, i.e. the lower and upper limits for $k_1$ and
$k_2$ in the integrals are always $\{-\pi/T_1, \pi/T_1\}$ and
$\{-\pi/T_2, \pi/T_2\}$ respectively, thus such lower and upper
limits will be omitted below.

When solution expansion (\ref{expansion}) is substituted into the
left hand side of Eq. (\ref{perturb_eq}), we get
\begin{equation}  \label{expand_semi}
\sum_{n=1}^\infty \int \!\!\! \int \phi_n(k_1, k_2) B_n(x, y; k_1,
k_2) \: dk_1 dk_2=\epsilon f(x, y) \psi(x, y),
\end{equation}
where $\phi_n(k_1, k_2)$ is defined as
\begin{equation}
\phi_n (k_1, k_2) \equiv \alpha_n (k_1, k_2) \left[\mu -\mu_n (k_1,
k_2)\right].
\end{equation}
Due to our localization assumption on the defect function $f(x, y)$,
the right hand side of Eq. (\ref{expand_semi}) is a 2D-localized
function, thus its Bloch-expansion coefficient $\phi_n(k_1, k_2)$ is
uniformly bounded in the Brillouin zone for all values of $n$ and
$\epsilon$. When solution expansion (\ref{expansion}) is further
substituted into the right hand side of Eq. (\ref{expand_semi}) and
orthogonality conditions (\ref{orth}) utilized, we find that
$\phi_n(k_1, k_2)$ satisfies the following integral equation
\begin{equation}
\phi_n (k_1, k_2)=\frac{\epsilon}{(2\pi)^2} \sum_{m=1}^\infty \int
\!\!\! \int \frac{\phi_m (\hat{k}_1, \hat{k}_2)}{\mu -\mu_m
(\hat{k}_1, \hat{k}_2)} W_{m,n} (k_1, k_2, \hat{k}_1, \hat{k}_2)
d\hat{k}_1d\hat{k}_2,   \label{integral}
\end{equation}
where function $W_{m,n}$ in the kernel is defined as
\begin{equation}
W_{m,n}(k_1, k_2, \hat{k}_1, \hat{k}_2)= \int_{-\infty}^\infty
\int_{-\infty}^\infty f(x, y) B_n^*(x, y; k_1, k_2)B_m(x, y;
\hat{k}_1, \hat{k}_2) dx dy.
\end{equation}
Again, since $f(x, y)$ is a 2D-localized function, $W_{m, n}(k_1,
k_2, \hat{k}_1, \hat{k}_2)$ is also uniformly bounded for all $(k_1,
k_2)$ and $(\hat{k}_1, \hat{k}_2)$ points in the Brillouin zone. The
fact of functions $\phi_n$ and $W_{m, n}$ being uniformly bounded is
important in the following calculations. Otherwise (when $f(x, y)$
is not 2D-localized, for instance), the results below would not be
valid.

At the edge point $\mu=\mu_c$, $\partial\mu_n /\partial k_1=\partial
\mu_n/
\partial k_2=0$. For simplicity, we also assume that
$\partial^2\mu_n /\partial k_1\partial k_2=0$ at this edge point ---
an assumption which is always satisfied for Eqs.
(\ref{eq:one})-(\ref{FD}) we considered in Sec. 2 due to symmetries
of its defective lattice. If $\partial^2\mu_n /\partial k_1\partial
k_2 \ne 0$ in some other problems, the analysis below only needs
minor modifications. Under the above assumption, the local
diffraction function near a $\Gamma$-symmetry edge point can be
expanded as
\begin{equation} \label{disp_expand}
\mu_n(k_1, k_2)=\mu_c+\gamma_1k_1^2+\gamma_2 k_2^2 +o(k_1^2, k_1k_2,
k_2^2),
\end{equation}
where
\begin{equation}
\gamma_1=\left. \frac{1}{2}\frac{\partial^2 \mu_n}{\partial
k_1^2}\right|_{(0, 0)}, \quad \gamma_2=\left.
\frac{1}{2}\frac{\partial^2 \mu_n}{\partial k_2^2} \right|_{(0, 0)}.
\end{equation}
Since $\mu=\mu_c$ is an edge point, which should certainly be a
local maximum or minimum point of the $n$-th diffraction surface,
clearly $\gamma_1$ and $\gamma_2$ must be of the same sign, i.e.
$\gamma_1\gamma_2>0$. The DM eigenvalue can be written as
\begin{equation}
\mu =\mu_c+\sigma h^2,     \label{mu_expand}
\end{equation}
where $\sigma=\pm 1$, and $0<h(\e)\ll 1$ when $\e \ll 1$. The
dependence of $h$ on $\e$ will be determined next.

When Eqs. (\ref{disp_expand}) and (\ref{mu_expand}) are substituted
into Eq. (\ref{integral}), we see that only a single term in the
summation with index $m=n$ makes $O(\phi_n)$ contribution. In this
term, the denominator $\mu-\mu_n (\hat{k}_1, \hat{k}_2)$ is very
small near the $\Gamma$ symmetry point (bifurcation point)
$(\hat{k}_1, \hat{k}_2)=(0, 0)$, which makes it $O(\phi_n)$ rather
than $O(\e\phi_n)$. The rest of the terms in the summation give
$O(\epsilon\phi_m)$ contributions, because the denominators
$\mu-\mu_m (\hat{k}_1, \hat{k}_2)$ in such terms are not small
anywhere in the first Brillouin zone. Thus
\begin{equation}
\phi_n (k_1, k_2)=\frac{\epsilon}{(2\pi)^2} \int \!\!\! \int
\frac{\phi_n (\hat{k}_1, \hat{k}_2)}{\mu -\mu_n (\hat{k}_1,
\hat{k}_2)} W_{n,n} (k_1, k_2, \hat{k}_1, \hat{k}_2)
d\hat{k}_1d\hat{k}_2 +O(\epsilon\phi_n).     \label{integral2}
\end{equation}
Since the first term on the right hand side of the above equation is
$O(\phi_n)$ rather than $O(\epsilon\phi_n)$, it can balance the left
hand side of that equation. This issue will be made more clear in
the calculations below. In order for the denominator in the integral
of Eq. (\ref{integral2}) not to vanish in the Brillouin zone, we
must require that
\begin{equation} \label{sigma}
\sigma = - \mbox{sgn}(\gamma_1)=- \mbox{sgn}(\gamma_2).
\end{equation}
This simply means that the DM eigenvalue $\mu$ must lie inside the
band gap as expected.

Now we substitute expressions (\ref{disp_expand}) and
(\ref{mu_expand}) into Eq. (\ref{integral2}), and can easily find
that
\begin{equation}
\phi_n (k_1, k_2)=\frac{\epsilon\sigma}{(2\pi)^2} \int \!\!\! \int
\frac{\phi_n (\hat{k}_1, \hat{k}_2)}{h^2 +
|\gamma_1|\hat{k}_1^2+|\gamma_2||\hat{k}_2^2} W_{n,n} (k_1, k_2,
\hat{k}_1, \hat{k}_2) d\hat{k}_1d\hat{k}_2 +O(\epsilon\phi_n).
\label{integral3}
\end{equation}
The above equation can be further simplified, up to error
$O(\epsilon\phi_n)$, as
\begin{equation}
\phi_n (k_1, k_2)=\frac{\epsilon\sigma}{(2\pi)^2} \phi_n (0,0)
W_{n,n} (k_1, k_2, 0,0) \int \!\!\! \int \frac{1}{h^2 +
|\gamma_1|\hat{k}_1^2+|\gamma_2||\hat{k}_2^2} d\hat{k}_1d\hat{k}_2
+O(\epsilon\phi_n). \label{integral4}
\end{equation}
To calculate the integral in the above equation, we introduce
variable scalings: $\tilde{k}_1=\sqrt{|\gamma_1|}\:\hat{k}_1,
\tilde{k}_2=\sqrt{|\gamma_2|}\:\hat{k}_2$. Then Eq.
(\ref{integral4}) becomes
\begin{equation}
\phi_n (k_1,
k_2)=\frac{\epsilon\sigma}{(2\pi)^2\sqrt{\gamma_1\gamma_2}} \phi_n
(0,0) W_{n,n} (k_1, k_2, 0,0) \int \!\!\! \int \frac{1}{h^2 +
\tilde{k}_1^2+\tilde{k}_2^2} d\tilde{k}_1d\tilde{k}_2
+O(\epsilon\phi_n), \label{integral5}
\end{equation}
where the integration is over the scaled Brillouin zone with
$|\tilde{k}_1| \le \pi\sqrt{|\gamma_1|}/T_1$ and $|\tilde{k}_2| \le
\pi\sqrt{|\gamma_2|}/T_2$. This integration region can be replaced
by a disk of radius $\sqrt{1-h^2}$ in the $(\tilde{k}_1,
\tilde{k}_2)$ plane, which causes error of $O(\epsilon\phi_n)$ to
Eq. (\ref{integral5}). Over this disk of radius $\sqrt{1-h^2}$, the
integral in Eq. (\ref{integral5}) can be easily calculated using
polar coordinates to be $-2\pi\ln h$, thus Eq. (\ref{integral5})
becomes
\begin{equation}
\phi_n (k_1, k_2)=-\frac{\epsilon\sigma\ln
h}{2\pi\sqrt{\gamma_1\gamma_2}} \phi_n (0,0) W_{n,n} (k_1, k_2, 0,0)
+O(\epsilon\phi_n). \label{integral6}
\end{equation}
Lastly, we take $k_1=k_2=0$ in the above equation. In order for it
to be consistent, we must have
\begin{equation}
\ln h=-\frac{2\pi \sigma \sqrt{\gamma_1\gamma_2}}{\epsilon
W_{n,n}(0,0,0,0)}+O(1).
\end{equation}
When this equation is substituted into (\ref{mu_expand}), we finally
get a formula for the DM eigenvalue $\mu$ as
\begin{equation} \label{mu_formula}
\mu=\mu_c+ \sigma C e^{-\beta/\e},
\end{equation}
where $\sigma$ is given by Eq. (\ref{sigma}),
\begin{equation} \label{beta}
\beta= \frac{4\pi \sigma \sqrt{\gamma_1\gamma_2}}{
W_{n,n}(0,0,0,0)},
\end{equation}
and $C$ is some positive constant. Obviously $\beta$ and $\epsilon$
must have the same sign, thus $\e$ and $\sigma W_{n,n}(0,0,0,0)$
must have the same sign. Since we have shown that $\sigma$ and
$\gamma_1$, $\gamma_2$ have opposite signs, we conclude that the
condition for defect-mode bifurcations from a $\Gamma$-symmetry edge
point is that
\begin{equation} \label{bifur_cond}
\mbox{sgn}\left[\e
W_{n,n}(0,0,0,0)\right]=-\mbox{sgn}(\gamma_1)=-\mbox{sgn}(\gamma_2).
\end{equation}
Under this condition, the DM eigenvalue $\mu$ bifurcated from the
edge point $\mu_c$ is given by formula (\ref{mu_formula}). Its
distance from the edge point, i.e. $\mu-\mu_c$, is exponentially
small with the defect strength $|\epsilon|$. This contrasts the 1D
case where such dependence is quadratic \cite{YangChendefect}. The
constant $C$ in formula (\ref{mu_formula}) is much more difficult to
calculate analytically, thus will not be pursued here.

If the edge point of DM bifurcations is not at a $\Gamma$ symmetry
point, the DM bifurcation condition (\ref{bifur_cond}) and the DM
eigenvalue formula (\ref{mu_formula}) still hold, except that
$\gamma_1, \gamma_2$ and $W_{n,n}$ in these formulas should now be
evaluated at the underlying edge point in the Brillouin zone.

Now we apply the above general results to the special system
(\ref{eq:two})-(\ref{eq:four}) we were considering in Sec. 2. When
$\e \ll 1$, this system can be written as
\begin{equation}
u_{xx}+u_{yy}+ \left[ \mu+V(x, y) \right] u =\epsilon f(x, y)u +
O(\e^2), \label{perturb_eq2}
\end{equation}
where $V(x, y)$ is given in Eq. (\ref{V}), and
\begin{equation}  \label{fxy}
f(x, y)=-\frac{E_0I_0\cos^2(x)\cos^2(y) F_D(x,
y)}{\left[1+I_0\cos^2(x)\cos^2(y)\right]^2}.
\end{equation}
In this case, $W(0, 0, 0, 0)$ is always negative, thus DM
bifurcation condition (\ref{bifur_cond}) reduces to
\begin{equation} \label{bifur_cond2}
\mbox{sgn}(\e)=\mbox{sgn}(\gamma_1)=\mbox{sgn}(\gamma_2).
\end{equation}
Thus, in an attractive defect, DMs bifurcate out from
normal-diffraction band edges (i.e. lower edges) of Figure
\ref{figure1}(a) where the diffraction coefficients are positive; in
a repulsive defect, DMs bifurcate out from anomalous-diffraction
band edges (i.e. upper edges) of Figure \ref{figure1}(a) where the
diffraction coefficients are negative. This analytical result is in
agreement with the heuristic argument in the end of Section 2.

It should be mentioned that at some band edges, two linearly
independent Bloch modes exist. Due to the symmetry of the lattice in
Eq. (\ref{eq:two}), these two Bloch modes are related as $B(x, y)$
and $B(y, x)$, where $B(x, y)\ne B(y, x)$. Under weak defects
(\ref{FD}), two different DMs of forms $u(x, y)$ and $u(y, x)$ but
with identical eigenvalues $\mu$ will bifurcate out from these two
Bloch modes of edge points respectively. This can be plainly seen
from the above analysis. Since these DMs have the same propagation
constant, their linear superposition remains a defect mode due to
the linear nature of Eq. (\ref{eq:four}). Such superpositions can
create more interesting DM patterns. This issue will be explored in
more detail in the next subsection.


\subsection{Numerical results of defect modes in localized defects of various depths}\label{section3}
The analytical results of the previous subsection hold under weak
localized defects. If the defect becomes strong (i.e. $|\e|$ not
small), the DM formula (\ref{mu_formula}) will be invalid. In this
subsection, we determine DMs in Eq. (\ref{eq:four}) numerically for
both weak and strong localized defects, and present various types of
DM solutions. In addition, in the case of weak defects, we will
compare numerical results with analytical ones in the previous
subsection, and show that they fully agree with each other. The
numerical method we will use is a power-conserving squared-operator
method developed in \cite{YangLakoba}.

\begin{figure}
\centering
\includegraphics[width=0.5\textwidth]{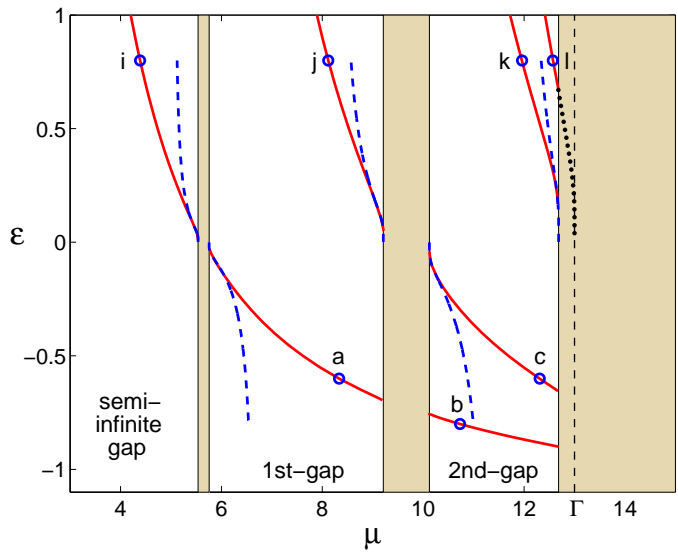}
\caption{Bifurcations of defect modes with the defect described by
Eq.~(\ref{eq:two}) at $E_0=15$ and $I_0=6$. Solid lines: numerical
results; dashed lines: analytical results. The shaded regions are
the Bloch bands. Profiles of DMs at the circled points in this
figure are displayed in Figs. \ref{figure3} and \ref{figure4}. }
\label{figure2}
\end{figure}

In these numerical computations, we fix $E_0=15$, $I_0=6$, and vary
the defect strength parameter $\epsilon$ from $-1$ to 1. We found a
number of DM branches which are plotted in Figure \ref{figure2}
(solid lines). First, we look at these numerical results under weak
defects ($|\e|\ll 1$). In this case, we see that DMs bifurcate out
from edges of every Bloch band. When the defect is attractive
($\e>0$), the bifurcation occurs at the left edge of each Bloch
band, while when the defect is repulsive ($\e<0$), the bifurcation
occurs at the right edge of each Bloch band. Recall that the left
and right edges of Bloch bands in Figure \ref{figure2} correspond to
the lower and upper edges of Bloch bands in Figure \ref{figure1}(a)
(where diffractions are normal and anomalous respectively), thus
these DM bifurcation results qualitatively agree with the
theoretical analysis in the previous subsection. We can further make
quantitative comparisons between numerical values of $\mu$ and the
theoretical formula (\ref{mu_formula}). This will be done in two
different ways. One way is that we have carefully examined the
numerical $\mu$ values for $|\e|\ll 1$, and found that they are
indeed well described by functions of the form (\ref{mu_formula}).
Data fitting of numerical values of $\mu$ into the form
(\ref{mu_formula}) gives $\beta$ values which are very close to the
theoretical values of (\ref{beta}). For instance, for the DM branch
in the semi-infinite bandgap in Figure \ref{figure2}, numerical data
fitting for $0<\e\ll 1$ gives $\beta_{num}=0.1289$
($C_{num}=0.4870$). The theoretical $\beta$ value, obtained from
Eqs. (\ref{beta}) and (\ref{fxy}), is $\beta_{anal}=0.1297$, which
agrees with $\beta_{num}$ very well. Another way of quantitative
comparison we have done is to plot the theoretical formula
(\ref{mu_formula}) alongside the numerical curves in Figure
\ref{figure2}. To do so, we first calculate the theoretical value
$\beta$ from Eq. (\ref{beta}) at each band edge. Regarding the
constant $C$ in the $\mu$ formula (\ref{mu_formula}), we do not have
a theoretical expression for it. To get around this problem, we fit
this single constant from the numerical values of $\mu$. The
theoretical formulas thus obtained at every Bloch-band edge are
plotted in Figure \ref{figure2} as dashed lines. Good quantitative
agreement between numerical and analytical values can be seen as
well.

As $|\e|$ increases, DM branches move away from band edges (toward
the left when $\e>0$ and toward the right when $\e<0$). Notice that
in attractive defects ($\e>0$), these branches stay inside their
respective bandgaps. But in repulsive defects ($\e<0$), DM branches
march to edges of higher Bloch bands and then re-appear in higher
bandgaps. For instance, the DM branch in the first bandgap reaches
the edge of the second Bloch band at $\e\approx -0.70$ and
re-appears in the second bandgap when $\e \approx -0.76$. These
behaviors resemble those in the 1D case (see Fig. 4 in
\cite{YangChendefect}). One difference between the 1D and 2D cases
seems to be, in 1D repulsive defects, when DM branches approach
edges of higher Bloch bands, the curves become tangential to the
vertical band-edge line. This indicates that these 1D defect modes
can not enter Bloch bands as embedded eigenvalues inside the
continuous spectrum. This fact is consistent with the statement
``(discrete) eigenvalues cannot lie in the continuous spectrum" in
\cite{Beketov} for locally perturbed periodic potentials in 1D
linear Schr\"odinger equations. In the present 2D repulsive defects,
on the other hand, the curves seem to intersect the vertical
band-edge lines non-tangentially. This may lead one to suspect that
these DM branches might enter Bloch bands and become embedded
discrete eigenvalues. This suspicion appears to be wrong however for
two reasons. First, the mathematical work by Kuchment and Vainberg
\cite{Kuchment} shows that for separable periodic potentials
perturbed by localized defects, embedded DMs can not exist. In our
theoretical model (\ref{eq:four}), the periodic potential is
non-separable, thus their result does not directly apply. But their
result strongly suggests that embedded DMs can not exist either in
our case. Second, our preliminary numerical computations support
this non-existence of embedded DMs. We find numerically that when a
DM branch enters Bloch bands in Figure \ref{figure2}, the DM mode
starts to develop oscillatory tails in the far field and becomes
non-localized (thus not being a DM anymore).

\begin{figure}[t]
\centering
\includegraphics[width=0.65\textwidth]{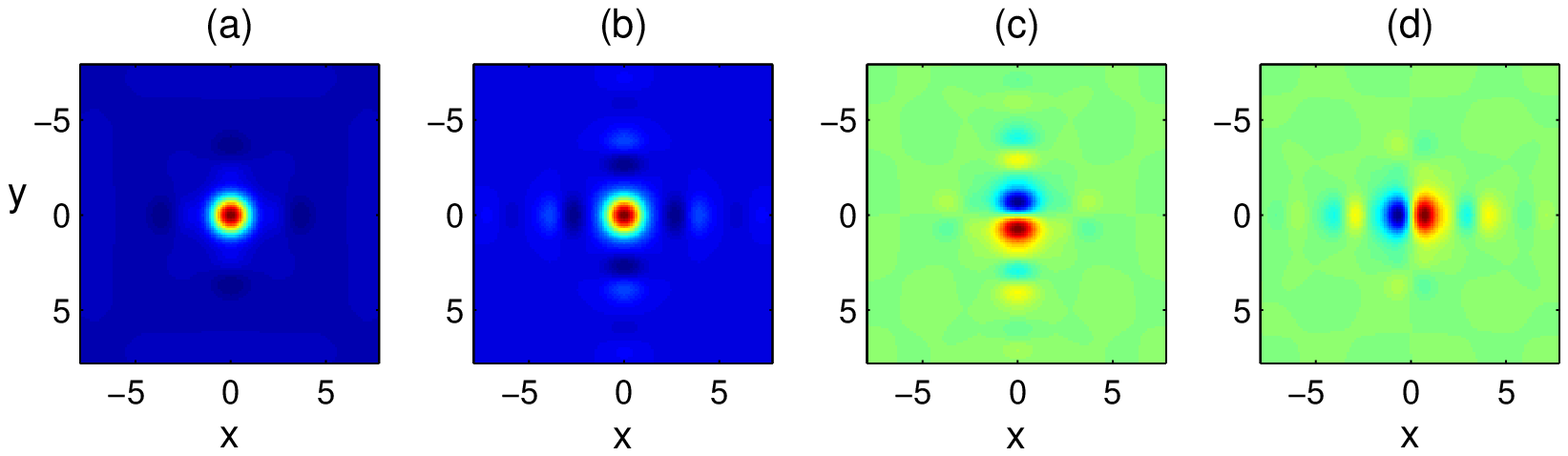}

\vspace{0.4cm}
\includegraphics[width=0.65\textwidth]{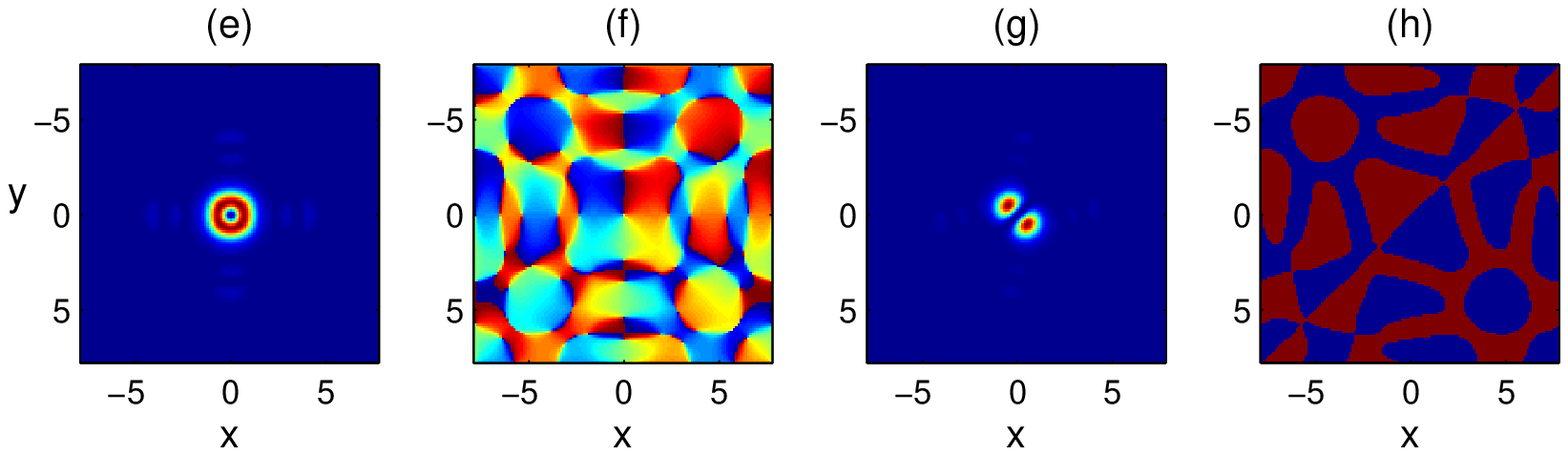}
\caption{Profiles of defect modes in repulsive defects in Figure
\ref{figure2}. (a)-(c): DMs at the circled points `a, b, c' in
Figure \ref{figure2}, with $(\e, \mu)$=$(-0.6, 8.33)$, $(-0.8,
10.73)$, and $(-0.6, 12.31)$, respectively; (d) the coexisting DM of
mode (c); (e) and (f): intensity and phase of the vortex mode
obtained by superimposing modes (c) and (d) with $\pi/2$ phase
delay, i.e. in the form of $u(x,y)+iu(y,x)$; (g) and (h): intensity
and phase of the diagonally-oriented dipole mode obtained by
superimposing modes (c) and (d) with no phase delay, i.e. in the
form of $u(x,y)+u(y,x)$.} \label{figure3}
\end{figure}

\begin{figure}[t]
\centering
\includegraphics[width=0.5\textwidth]{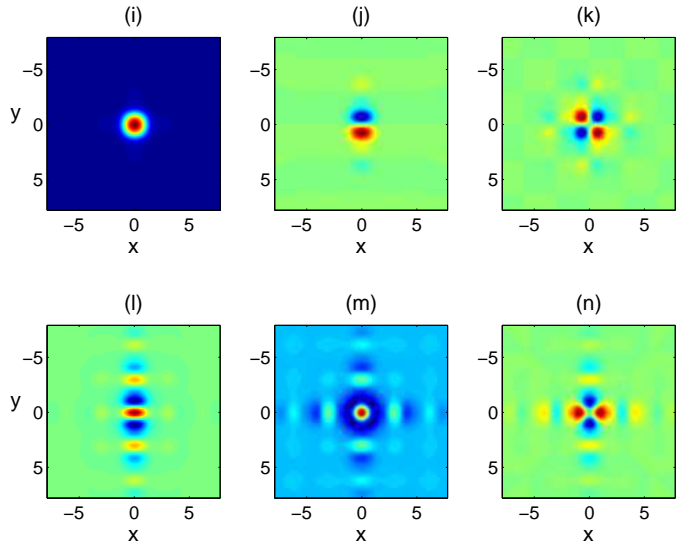}
\caption{Profiles of defect modes in attractive defects in Figure
\ref{figure2}. (i)-(l): DMs at the circled points `i, j, k, l' in
Figure \ref{figure2}, with $(\e, \mu)$=(0.8, 4.39), (0.8, 8.12),
(0.8, 11.96), and (0.8, 12.57), respectively. (m) and (n): DMs
obtained by superimposing mode (l) and its coexisting mode with zero
and $\pi$ phase delays, i.e. in the form of $u(x,y)+u(y,x)$ and
$u(x,y)-u(y,x)$ respectively.} \label{figure4}
\end{figure}

Now we examine profiles of defect modes on DM branches in Figure
\ref{figure2}. For this purpose, we select one representative point
from each DM branch, mark them by circles, and label them by letters
in Figure \ref{figure2}. DM profiles at these marked points are
displayed in Figs. \ref{figure3} and \ref{figure4}. The letter
labels for these defect modes are identical to those for the marked
points on DM branches in Figure \ref{figure2}. First, we look at
Figure \ref{figure3}, which shows DM profiles in repulsive defects
$(\e <0)$. We see that DMs at points `a, b' of Figure \ref{figure2}
are symmetric in both $x$ and $y$, with a dominant hump at the
defect site, and satisfy the relation $u(x, y)=u(y, x)$. These DMs
are the simplest in their respective bandgaps, thus we will call
them fundamental DMs. Notice that these fundamental DMs are
sign-indefinite, i.e. they have nodes where the intensities are
zero, because they do not lie in the semi-infinite bandgap. Although
DMs in Figure \ref{figure3}(a, b) look similar, differences between
them (mainly in their tail oscillations) do exist due to their
residing in different bandgaps. These differences are qualitatively
similar to the 1D case, which has been carefully examined before
(see Fig. 8 (b, d) in \cite{YangChendefect} and Fig. 4 (b, d) in
\cite{Chen_OE}). The DM branch of point `c' is more interesting. At
each point on this branch, there are two linearly independent DMs,
the reason being that this DM branch bifurcates from the right edge
of the second Bloch band where two linearly independent Bloch modes
exist (this band edge is a $X$ symmetry point, see Fig. 1). These
two DMs are shown in Figure \ref{figure3}(c, d). One of them is
symmetric in $x$ and anti-symmetric in $y$, while the other is
opposite. They are related as $u(x, y)$ and $u(y, x)$ with $u(x,
y)\ne u(y, x)$. These DMs are dipole-like. However, these dipoles
are largely confined inside the defect site. They are closely
related to certain single-Bloch-mode solitons reported in
\cite{Shi}.

Since the DM branch of point `c' admits two linearly independent
defect modes, their arbitrary linear superposition would remain a
defect mode since Eq. (\ref{eq:four}) is linear. Such linear
superpositions could lead to new interesting DM structures. For
instance, if the two DMs in Figure \ref{figure3}(c, d) are
superimposed with $\pi/2$ phase delay, i.e. in the form of $u(x,
y)+iu(y, x)$, we get a vortex-type DM whose intensity and phase
structures are shown in Figure \ref{figure3}(e, f). This vortex DM
is strongly confined in the defect site and looks like a familiar
vortex-ring. It is qualitatively similar to vortex-cell solitons in
periodic lattices under defocusing nonlinearity as reported in
\cite{Shi}, as well as linear vortex arrays in defected lattices as
observed in \cite{ChenYangPRL}. However, it is quite different from
the gap vortex solitons as observed in \cite{Segev_highervortex},
where the vortex beam itself creates an attractive defect with
focusing nonlinearity, while the vortex DM here is supported by a
repulsive defect. If the two DMs in Figure \ref{figure3}(c, d) are
directly superimposed without phase delays, i.e. in the form of
$u(x, y)+u(y, x)$, we get a dipole-type DM whose intensity and phase
profiles are shown in Figure \ref{figure3}(g, h). This dipole DM is
largely confined in the defect site and resembles the DM of Figure
\ref{figure3}(c, d), except that its orientation is along the
diagonal direction instead. This dipole DM is qualitatively similar
to dipole-cell solitons in uniform lattices under defocusing
nonlinearity as reported in \cite{Shi}. If the two DMs in Figure
\ref{figure3}(c, d) are superimposed with $\pi$ phase delay, i.e. in
the form of $u(x, y)-u(y, x)$, the resulting DM would also be
dipole-like but aligned along the other diagonal axis. Such a DM is
structurally the same as the one shown in Figure \ref{figure3}(g, h)
in view of the symmetries of our defected lattice.

Now we examine DMs in attractive defects, which are shown in Figure
\ref{figure4}. At point `i' in the semi-infinite bandgap, the DM is
bell-shaped and is strongly confined inside the defect site [see
Figure \ref{figure4}(i)]. This mode is guided by the total internal
reflection mechanism, which is different from the repeated Bragg
reflection mechanism for DMs in higher bandgaps. On the DM branch of
point `j', there are two linearly independent DMs which are related
as $u(x, y)$ and $u(y, x)$. One of them is shown in Figure
\ref{figure4}(j). This mode is dipole-like and resembles the one in
Figure \ref{figure3}(c). Linear superpositions of this mode $u(x,
y)$ with its coexisting mode $u(y, x)$ would generate vortex- and
dipole-like DMs similar to those in Figure \ref{figure3}(e, f, g,
h). In particular, the vortex DM at this point `j' would be
qualitatively similar to the gap vortex soliton as observed in
\cite{Segev_highervortex}. On the DM branch of point `k', a single
DM exists. This is because the left edge of the third Bloch band
where this branch of DM bifurcates from is located at the $M$
symmetry point of the Brillouin zone (see Fig. 1) and admits only a
single Bloch mode. The DM at point `k' is displayed in Figure
\ref{figure4}(k). This mode is largely confined at the defect site
and is quadrupole-like. On the DM branch of point `l', two linearly
independent DMs exist which are related as $u(x, y)$ and $u(y, x)$.
One of them is shown in Figure \ref{figure4}(l). This mode is
symmetric in both $x$ and $y$ directions and is tripolar-like (with
three dominant humps). Unlike DMs at points `c, j', this DM,
superimposed with its coexisting mode with $\pi/2$ phase delay would
{\it not} generate vortex modes, since this DM has non-zero
intensity at the center of the defect. However, their superpositions
with zero and $\pi$ phase delays, i.e. in the forms of $u(x, y)+u(y,
x)$ and $u(x, y)-u(y, x)$ would lead to two structurally different
defect modes, which are shown in Figure \ref{figure4}(m, n). The DM
in Figure \ref{figure4}(m) has a dominant hump in the center of the
defect, surrounded by a negative ring, and with weaker satellite
humps further out. The DM in Figure \ref{figure4}(n) is
quadrupole-like, but oriented differently from the quadrupole-like
DM in Figure \ref{figure4}(k). These DMs resemble the spike-array
solitons and quadrupole-array solitons in uniform lattices under
focusing nonlinearity as reported in \cite{Shi}.

Most of the DM branches in Figure \ref{figure2} bifurcate from edges
of Bloch bands. Even the branch of point `b' in the second bandgap
of Figure \ref{figure2} can be traced to the DM bifurcation from the
right edge of the first Bloch band. But there are exceptions. One
example is the DM branch of point `l' in the upper right corner of
the second bandgap in Figure \ref{figure2} (we will call it the `l'
branch below). This `l' branch does {\it not} bifurcate from any
Bloch-band edge. Careful examinations show that DMs on this branch
closely resemble Bloch modes at the lowest $\Gamma$-symmetry point
in the third Bloch band (see Figure \ref{figure1}). Thus this `l'
branch should be considered as bifurcating from that lowest
$\Gamma$-symmetry point when $\e\ll 1$. However, we see from Figure
\ref{figure1} that this lowest $\Gamma$-symmetry point is {\it not}
an edge point of the third Bloch band. Even though it is a {\it
local} edge (minimum) point of diffraction surfaces in the Brillouin
zone, it still lies {\it inside} the third Bloch band. Let us call
this type of points ``quasi-edge points" of Bloch bands. Then the
`l' branch of DMs bifurcates from a quasi-edge point, not a true
edge point, of a Bloch band! To illustrate this fact, we connected
the `l' branch to this quasi-edge $\Gamma$-symmetry point (at
$\e=0$) by dotted lines through a fitted function of the form
(\ref{mu_formula}) in Figure \ref{figure2}. This dotted line lies
inside the third Bloch band. One question we can ask here is: what
is the nature of this dotted line? Since it is inside the Bloch
band, the mathematical results by Kuchment and Vainberg
\cite{Kuchment} suggest that it can not be a branch of embedded DMs.
On the other hand, the `l' DM branch in the second bandgap does
bifurcate out along this route. Then how does the `l' branch
bifurcate from the quasi-edge point along this route? This question
awaits further investigation. We note by passing that the $\Gamma$
and $M$ points inside the second Bloch band are also quasi-edge
points (see Figure \ref{figure1}(a)). Additional DM branches may
bifurcate from such quasi-edge points as well.

\section{Dependence of defect modes on the applied dc field $E_0$   \label{DC_dependence}}
In the previous section, we investigated the bifurcation of DMs as
the local-defect strength parameter $\e$ varies. In experiments on
photorefractive crystals, the applied dc field $E_0$ can be adjusted
in a wide range much more easily than the defect strength. So in
this section, we investigate how DMs are affected by the value of
$E_0$ in Eqs. (\ref{eq:one})-(\ref{eq:two}) under localized defects
of (\ref{FD}). When $E_0$ changes, so does the lattice potential. We
consider both attractive and repulsive defects at fixed values of
$\epsilon=\pm0.9$ (the reason for these choices of $\e$ values will
be given below). For other fixed $\e$ values, the dependence of DMs
on $E_0$ is expected to be qualitatively the same. The reader is
reminded that throughout this section, the $I_0$ value is fixed at
$I_0=6$.

\subsection{The case of repulsive defects}

\begin{figure}
\centering
\includegraphics[width=0.25\textwidth]{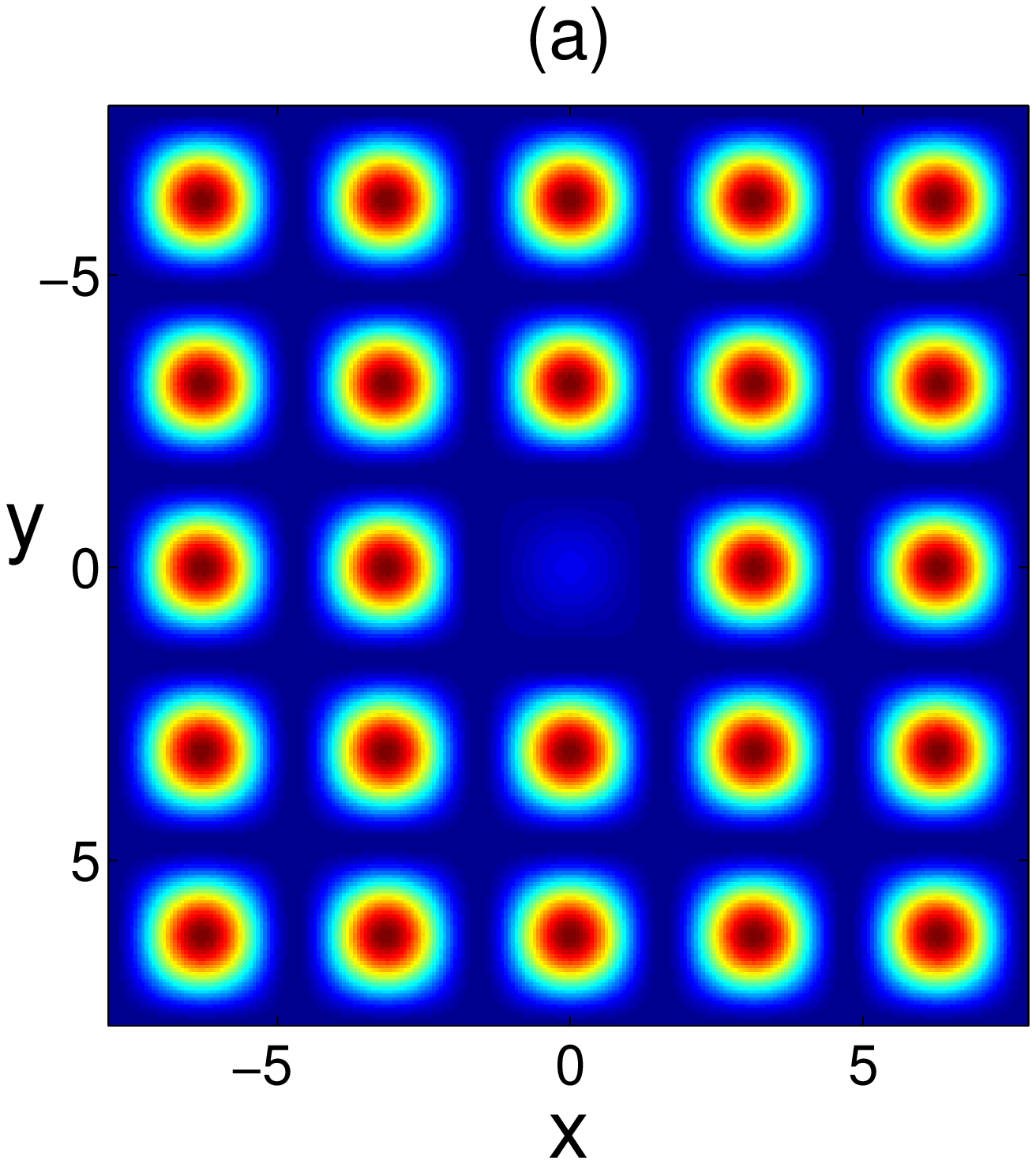}
\hspace{1cm}
\includegraphics[width=0.4\textwidth]{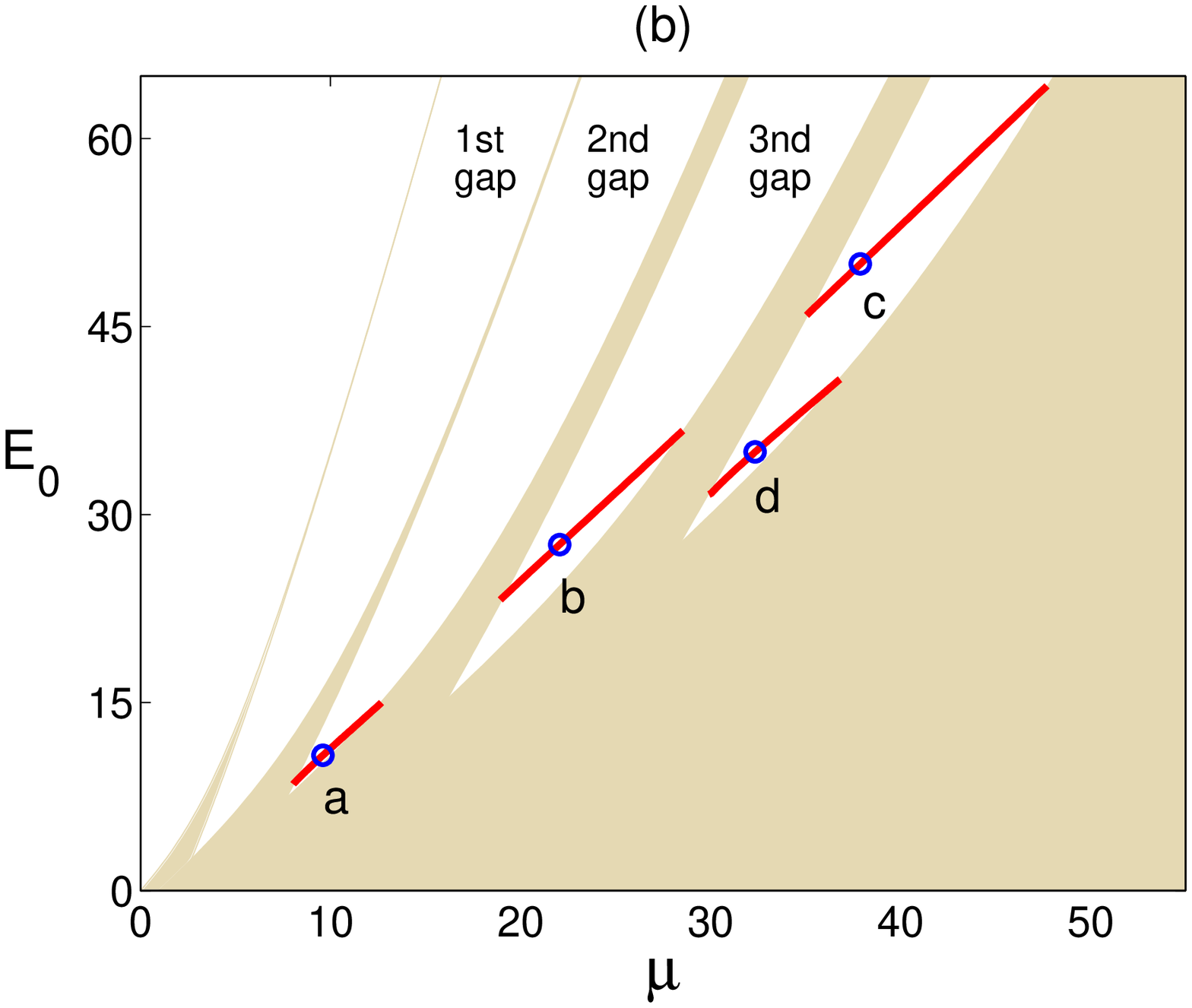}
\caption{(a) Profile of the lattice $I_L(x,y)$ in Eq.~(\ref{eq:two})
with a repulsive localized defect of (\ref{FD}) and $\epsilon=-0.9$
($I_0=6$). (b) The corresponding DM branches in the $(\mu, E_0)$
parameter space. Shaded: Bloch bands. DMs at the circled points are
displayed in Figure \ref{figure6}.} \label{figure5}
\end{figure}

In 2D photonic lattices with single-site repulsive defects we
experimentally created in \cite{ChenYangPRL}, $\e \approx -1$. At
this value of $\e$, however, we found that DMs may exist in high
bandgaps, but not in low bandgaps (see Figure \ref{figure2} where
$E_0=15$). Thus for presentational purposes, we take $\epsilon=-0.9$
in our computations below. The corresponding lattice profile of
$I_L(x,y)$ is shown in Figure \ref{figure5}(a), which still closely
resembles the defects in our previous experiments
\cite{ChenYangPRL}.

\begin{figure}
\centering
\includegraphics[width=0.5\textwidth]{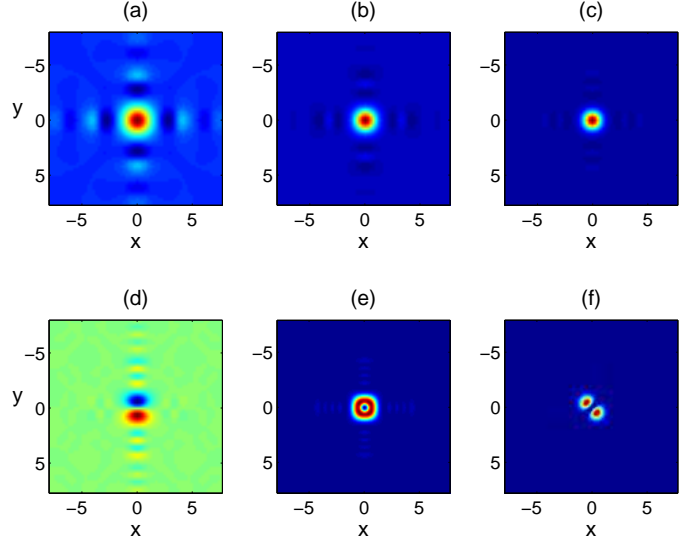}
\caption{(a)-(d): Defect modes supported by the repulsive localized
defect of Figure \ref{figure5}(a) at the circled points `a, b, c, d'
of Figure \ref{figure5}(b), with $(E_0, \mu)$=(10.8, 9.61), (27.6,
22.06), (50.0, 37.88), and (35.0, 32.34), respectively. (e) and (f):
Intensity profiles of the vortex and dipole modes obtained by
superposing mode (d) and its coexisting mode with $\pi/2$ and zero
phase delays. } \label{figure6}
\end{figure}

For this defect, we found different types of defect modes at various
values of $E_0$. The results are summarized in Figs.
\ref{figure5}(b) and \ref{figure6}. In Figure \ref{figure5}(b), the
dependence of DM eigenvalues $\mu$ on the dc field $E_0$ is
displayed. As can be seen, when $E_0$ increases from zero, a DM
branch appears in the second bandgap at $E_0\approx8.5$. When $E_0$
increases further, this branch moves toward the third Bloch band and
disappears at $E_0\approx15$. But when $E_0$ increases to
$E_0\approx23.4$, another DM branch appears in the third bandgap.
This branch moves toward the fourth Bloch band and disappears at
$E_0\approx36.7$. In the fourth gap, there are two DM branches. The
upper branch exists when $45.9<E_0<64.2$, and the lower branch
exists when $31.6<E_0<40.8$. As $E_0$ increases even further, DM
branches will continue to migrate from lower bandgaps to higher
ones. These behaviors are qualitatively similar to the 1D case (see
Fig. 7 in \cite{YangChendefect}). Some examples of DMs marked by
circles in Figure \ref{figure5} are displayed in Figure
\ref{figure6}. We can see that DMs at points `a, b, c' are
``fundamental" DMs. At point `d', there are two linearly independent
DMs related by $u(x, y)$ and $u(y, x)$, one of which is displayed in
Figure \ref{figure6}(d). These DMs are dipole-like, similar to the
`c' branch in Figure \ref{figure2}. Linear superpositions of these
DMs with $\pi/2$ and $0$ phase delays would generate a vortex DM and
a diagonally-oriented dipole DM, whose amplitude profiles are shown
in Figure \ref{figure6} (e) and (f) (their phase fields are similar
to those in Figure \ref{figure3}(f, h) and thus not shown here).

\subsection{The case of attractive defects}
\begin{figure}
\centering
\includegraphics[width=0.25\textwidth]{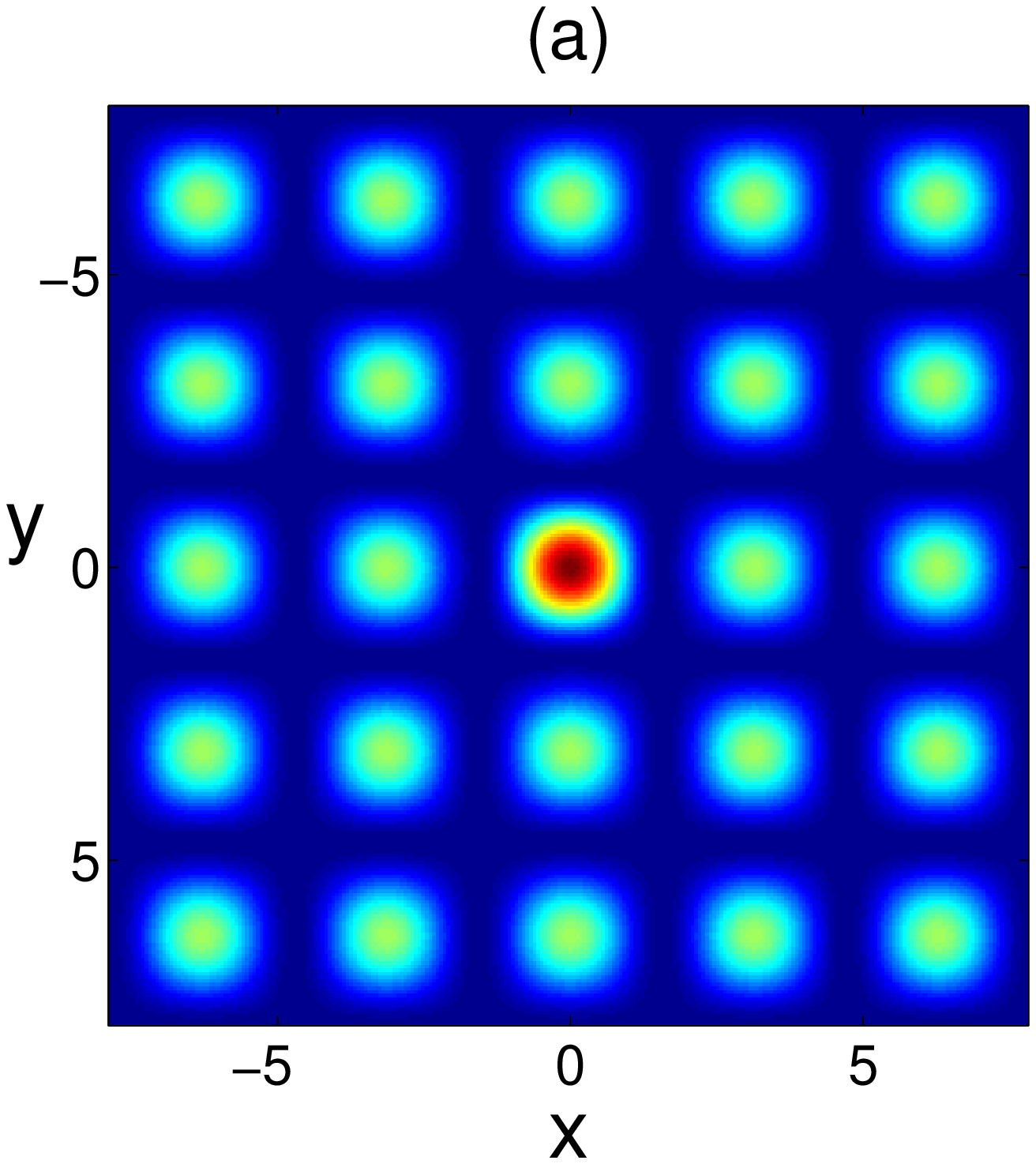}
\hspace{1cm}
\includegraphics[width=0.4\textwidth]{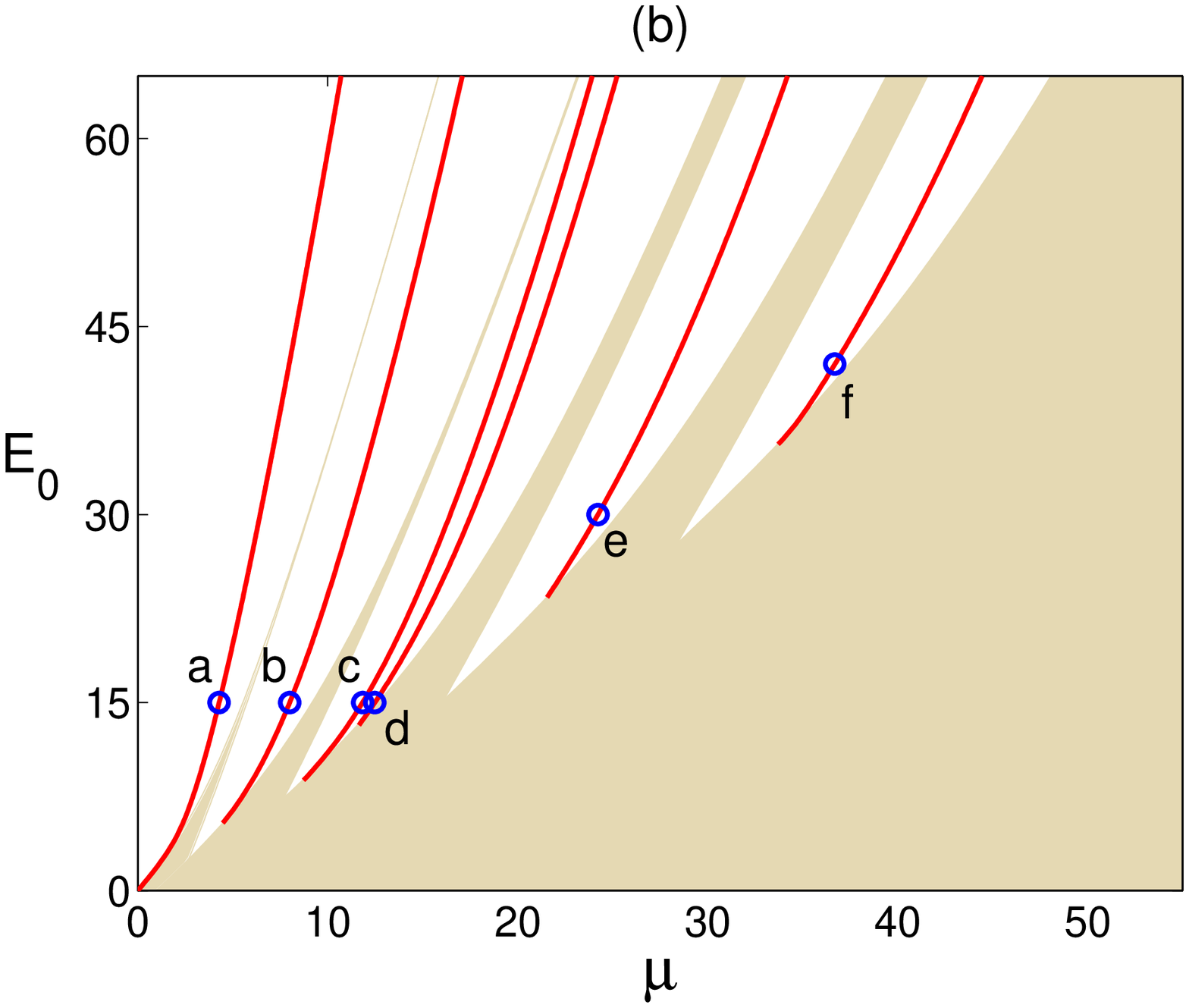}
\caption{(a) Profile of the lattice $I_L(x,y)$ in Eq.~(\ref{eq:two})
with an attractive localized defect of (\ref{FD}) and $\epsilon=0.9$
($I_0=6$). (b) The corresponding DM branches in the $(\mu, E_0)$
parameter space. Shaded: Bloch bands. DMs at the circled points are
displayed in Figure \ref{figure8}. } \label{figure7}
\end{figure}

\begin{figure}
\centering
\includegraphics[width=0.5\textwidth]{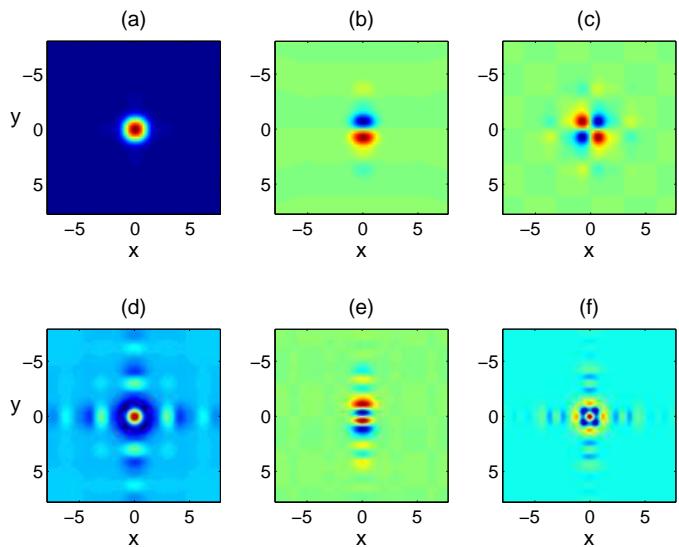}
\caption{Defect modes supported by attractive localized defects of
Figure \ref{figure7}(a) at the circled points `a, b, c, d, e, f' of
Figure \ref{figure7}(b), with $(E_0, \mu)$=(15, 4.29), (15, 8.00),
(15, 11.84), (15, 12.48), (30, 24.23) and (42, 36.69), respectively.
} \label{figure8}
\end{figure}

For attractive defects, we choose $\epsilon=0.9$ following the above
choice of $\epsilon=-0.9$ for repulsive defects. The lattice profile
at this $\epsilon$ value is shown in Figure \ref{figure7}(a). The
dependence of DM eigenvalues $\mu$ on the dc field $E_0$ is shown in
Figure \ref{figure7}(b). Unlike DMs in repulsive defects, DMs in
this attractive defect exist in all bandgaps (including the
semi-infinite and first bandgaps). In addition, DM branches here
stay in their respective bandgaps as $E_0$ increases, contrasting
the repulsive case. DM profiles at marked points in Figure
\ref{figure7}(b) are displayed in Figure \ref{figure8}. The DM in
Figure \ref{figure8}(a) is bell-shaped and all positive, and it is
the fundamental DM in the semi-infinite bandgap. This DM is similar
to the one on the `i' branch of Figure \ref{figure2}, and is guided
by the total internal reflection mechanism. The DM in the first
bandgap is dipole-like and is displayed in Figure \ref{figure8}(b).
This DM branch is similar to the `j' branch of Figure \ref{figure2}.
On this branch, there is another coexisting DM which is a
$90^\circ$-rotation of Figure \ref{figure8}(b). Linear
superpositions of these two DMs could generate vortex-like and
diagonally-oriented dipole-like DMs, as Figure \ref{figure3}(e, f,
g, h) shows. In the second bandgap, there are two DM branches: the
`c' branch and the `d' branch. DMs on the `c' branch are
quadrupole-like, and this branch is similar to the `k' branch of
Figure \ref{figure2}. On this branch, there is a single linearly
independent defect mode. The `d' branch is similar to the `l' branch
of Figure \ref{figure2}. On this branch, there are two linearly
independent DMs which are tripolar-like and orthogonal to each other
[see Figure \ref{figure4}(l)]. A superposition of these DMs with
zero phase delay creates a hump-ring-type structure which is shown
in Figure \ref{figure8}(d) [see also Figure \ref{figure4}(m)]. A
different superposition would produce a quadrupole-type structure as
the one shown in Figure \ref{figure4}(n). In the third and fourth
bandgaps, there are additional DM branches. These branches exist at
$E_0$ values higher than 15, thus have no counterparts in Figure
\ref{figure2} (where $E_0=15$). Profiles of these DMs are more
exotic as can be seen in Figure \ref{figure8} (e) and (f).

\section{Defect modes in non-localized defects} \label{non-localized}
In previous sections, we exclusively dealt with defect modes in
2D-localized defects. We found that DM eigenvalues bifurcate out
from Bloch-band edges exponentially with the defect strength $\e$
[see Eq. (\ref{mu_formula})]. In addition, DMs can not be embedded
inside Bloch bands (i.e. the continuous spectrum). In this section,
we briefly discuss DMs in non-localized defects, and point out two
significant differences between DMs in localized and non-localized
defects. One is that in non-localized defects, DM eigenvalues can
bifurcate out from edges of the continuous spectrum algebraically,
not exponentially, with the defect strength $\e$. The other one is
that in non-localized defects, DMs {\it can} be embedded inside the
continuous spectrum as embedded eigenmodes.

\begin{figure}
\centering
\includegraphics[width=0.25\textwidth]{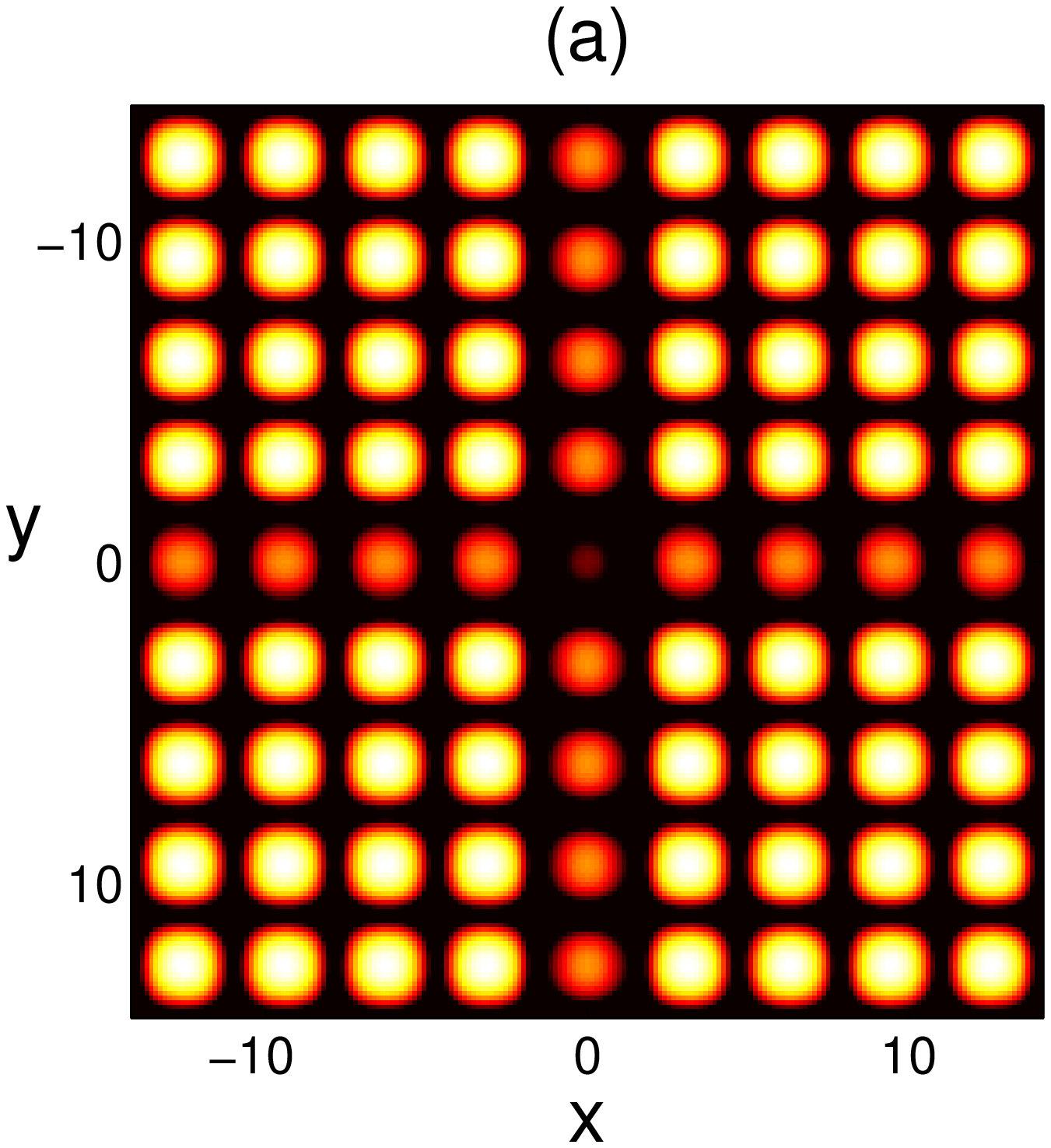}
\hspace{1cm}
\includegraphics[width=0.4\textwidth]{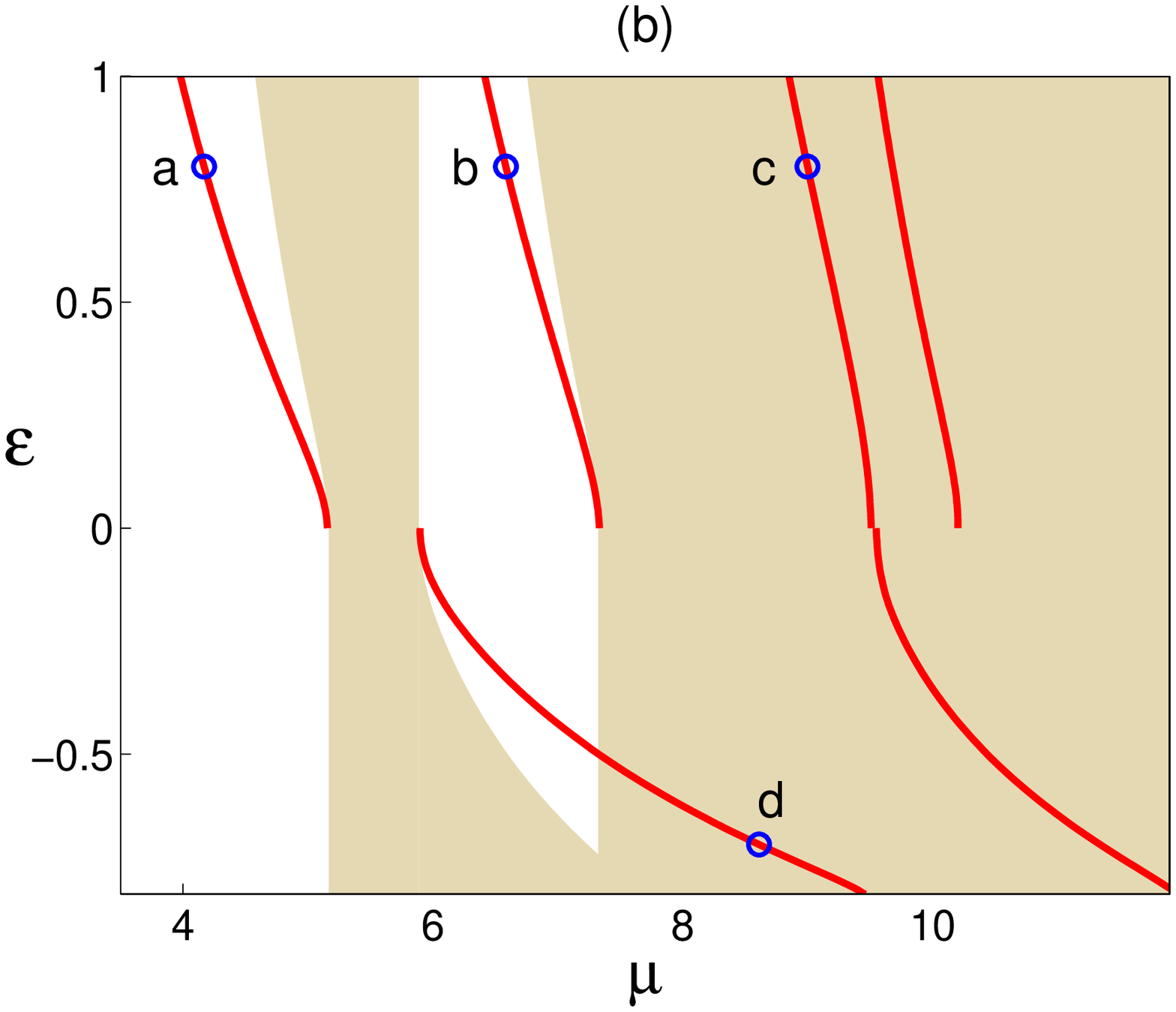}
\caption{(a) Profile of a 2D potential with non-localized defects
$V_D(x)+V_D(y)$ in Eqs.~(\ref{sep_defect})-(\ref{VD}) with $E_0=6$,
$I_0=3$, and $\epsilon=-0.7$. (b) DM branches supported by this
non-localized defect in the $(\mu, \e)$ parameter space. Shaded: the
continuous spectrum. DMs at the circled points are displayed in
Figure \ref{figure10}.} \label{figure9}
\end{figure}

For simplicity, we consider the following linear Schr\"odinger
equation with {\it separable} non-localized defects,
\begin{equation} \label{sep_defect}
u_{xx}+u_{yy}+\left\{V_D(x)+V_D(y)\right\}u=-\mu u,
\end{equation}
where function $V_D$ is a one-dimensional defected periodic
potential of the form
\begin{equation} \label{VD}
V_D(x)=-\frac{E_0}{1+I_0 \cos^2(x) \left\{1+\e F_D(x)\right\}},
\end{equation}
$F_D(x)=\mbox{exp}(-x^8/128)$ is a defect function as has been used
before \cite{YangChendefect}, and $\e$ is the defect strength. We
chose this separable defect in Eq. (\ref{sep_defect}) because all
its eigenmodes (both discrete and continuous) can be obtained
analytically (see below). The non-localized nature of this separable
defect can be visually seen in Figure \ref{figure9}(a), where the
defected potential $V_D(x)+V_D(y)$ for $I_0=3, E_0=6$, and $\e=-0.7$
is displayed. Note that this potential plotted here differs from the
usual concept of quantum-mechanics potentials by a sign, and it
resembles the refractive-index function in optics.
We see that along the $x$ and $y$ axes in Figure \ref{figure9}(a),
the defect extends to infinity (thus non-localized). The above $I_0$
and $E_0$ parameters were chosen because they have been used before
in the 1D defect-mode analysis of \cite{YangChendefect}, and such 1D
results will be used to construct the eigenmodes of the 2D defect
equation (\ref{sep_defect}) below.

Since the potential in Eq. (\ref{sep_defect}) is separable, the
eigenmodes of this equation can be split into the following form:
\begin{equation} \label{split}
u(x, y)=u_a(x)u_b(y),  \quad \mu=\mu_a+\mu_b,
\end{equation}
where $u_a, u_b, \mu_a, \mu_b$ satisfy the following one-dimensional
eigenvalue equation:
\begin{equation} \label{ua}
u_{a,xx}+V_D(x)u_a=-\mu_a u_a,
\end{equation}
\begin{equation}
u_{b,yy}+V_D(y)u_b=-\mu_b u_b.
\end{equation}
These two 1D equations for $u_a$ and $u_b$ are identical to the 1D
defect equation we studied comprehensively in \cite{YangChendefect}.
Using the splitting (\ref{split}) and the 1D results in
\cite{YangChendefect}, we can construct the entire discrete and
continuous spectra of Eq. (\ref{sep_defect}).

Before we construct the spectra of Eq. (\ref{sep_defect}), we need
to clarify the definitions of discrete and continuous eigenvalues of
this 2D equation. Here an eigenvalue is called discrete if its
eigenfunction is square-integrable (thus localized along all
directions in the $(x, y)$ plane). Otherwise it is called
continuous. Note that an eigenfunction which is localized along one
direction (say $x$ axis) but non-localized along another direction
(say $y$ axis) corresponds to a continuous, not discrete,
eigenvalue.

Now we construct the spectra of Eq. (\ref{sep_defect}) for a
specific example with $E_0=6, I_0=3$ and $\e=0.8$. At these
parameter values, the discrete eigenvalues and continuous-spectrum
intervals (1D Bloch bands) of the 1D eigenvalue problem (\ref{ua})
are (see \cite{YangChendefect})
\begin{equation}
\{\lambda_1, \lambda_2, \lambda_3, \dots\}= \{2.0847, \; 4.5002, \;
7.5951, \dots, \},
\end{equation}
\begin{equation}
\{[I_1, I_2],\, [I_3, I_4], \, [I_5, I_6], \dots\}= \{[2.5781, \;
2.9493], \; [4.7553, \; 6.6010], \; [7.6250, \; 11.8775], \dots \}.
\end{equation}
Using the relation (\ref{split}), we find that the discrete
eigenvalues and continuous-spectrum intervals of the 2D eigenvalue
problem (\ref{sep_defect}) are
\begin{equation}
\{\mu_1, \mu_2, \mu_3, \mu_4, \dots\}= \{2\lambda_1, \;
\lambda_1+\lambda_2, \; 2\lambda_2, \; \lambda_1+\lambda_3, \dots\}
= \{4.1694, \; 6.5849, \; 9.0004, \; 9.6798, \dots, \},
\end{equation}
and
\begin{equation}
\{\mu_{\mbox{continuum}}\}=\{[\lambda_1+I_1, \: 2I_2], \;
[\lambda_1+I_3, \: \infty) \} = \{[4.6628, \; 5.8986], \; [6.8400,
\;  \infty)\}.
\end{equation}
Note that at the left edges of the two continuous-spectrum bands
($\mu=4.6628$ and 6.8400), the eigenfunctions are non-localized
along one direction but localized along its orthogonal direction,
thus they are not the usual 2D Bloch modes (which would have been
non-localized along all directions). This is why for Eq.
(\ref{sep_defect}), we do not call these continuous-spectrum bands
as Bloch bands.

Repeating the same calculations to other $\e$ values, we have
constructed the whole spectra of Eq. (\ref{sep_defect}) in the
$(\mu, \e)$ plane for $I_0=3$ and $E_0=6$. The results are displayed
in Figure \ref{figure9}(b). Here solid curves show branches of DMs
(discrete eigenvalues), and shaded regions are the continuous
spectrum. This figure shows two significant features which are
distinctively different from those in localized defects. One is that
several DM branches (such as the `c' and `d' branches) are either
partially or completely embedded inside the continuous spectrum.
This means that in non-localized defects, embedded DMs inside the
continuous spectrum do exist. This contrasts the localized-defect
case, where the mathematical results of Kuchment and Vainberg
\cite{Kuchment} and our numerics show non-existence of embedded DMs.
Another feature of Figure \ref{figure9}(b) is on the quantitative
behavior of DM bifurcations from edges of the continuous spectrum at
small values of the defect strength $\e$. We have shown before that
in the 1D case, DMs bifurcate out from Bloch-band edges
quadratically with $\e$ \cite{YangChendefect}. For the 2D problem
(\ref{sep_defect}), using the relation (\ref{split}), we can readily
see that when DMs bifurcate out from edges of the continuous
spectrum (see the `a, b, d' branches for instance), the distance
between DM eigenvalues and the continuum edges also depends on $\e$
quadratically. This contrasts the localized-defect case, where we
have shown in section \ref{DMbifurcation} that DMs bifurcate out
from Bloch-band edges exponentially.

\begin{figure}[t]
\centering
\includegraphics[width=0.65\textwidth]{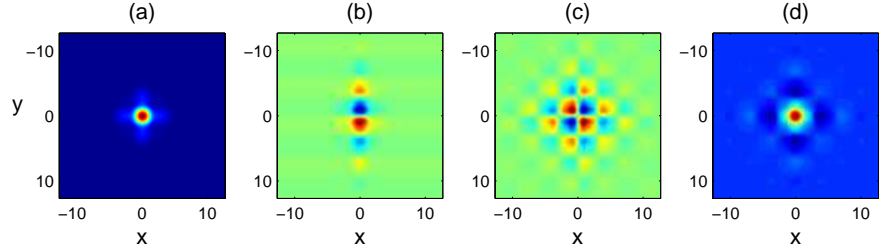}
\caption{Defect modes in non-localized defects
(\ref{sep_defect})-(\ref{VD}) at the circled points `a, b, c, d' in
Figure \ref{figure9}(b), with $(\e, \mu)$=(0.8, 4.17), (0.8,6.59),
(0.8, 9.00), and $(-0.7,8.61)$ respectively. } \label{figure10}
\end{figure}

Even though the defect in Eq. (\ref{sep_defect}) is non-localized
rather than localized, the DMs it admits actually are quite similar
to those in localized defects. To demonstrate, we picked four
representative points on the DM branches of Figure \ref{figure9}(b).
These points are marked by circles and labeled by letters `a, b, c,
d' respectively. Profiles of DMs at these four points are displayed
in Figure \ref{figure10}. Comparing these DMs with those of
localized defects in Figs. \ref{figure3} and \ref{figure4}, we
easily see that they are quite similar. In particular, the `a, b, c,
d' branches in Figure \ref{figure9}(b) resemble the `i, j, k, a'
branches in Figure \ref{figure2}, as DMs on the corresponding
branches are much alike. Notice that DMs in the non-localized defect
case are a little more spread out than their counterparts in the
local-defect case (compare Figs. \ref{figure4}(k) and
\ref{figure10}(c), for instance). This is simply because in the
non-localized defect, we chose $E_0=6$, while in the localized
defect, we chose a much higher value of $E_0=15$. Higher $E_0$
values induce deeper potential variations, which facilitates better
confinement of DMs. We need to point out that, even though the DMs
at points `c, d' are embedded inside the continuous spectrum, their
eigenfunctions are perfectly 2D-localized and square-integrable (see
Figure \ref{figure10}(d) in particular). Thus, the embedded nature
of a defect mode does not necessarily cause it to spread out more.

\section{Summary}
In summary, we have thoroughly investigated defect modes in
two-dimensional photonic lattices with localized or non-localized
defects. When the defect is localized and weak, we analytically
determined defect-mode eigenvalues bifurcated from edges of Bloch
bands. We found that in an attractive (repulsive) defect, defect
modes bifurcate out from Bloch-band edges with normal (anomalous)
diffraction coefficients. Furthermore, distances between defect-mode
eigenvalues and Bloch-band edges are exponentially small functions
of the defect strength, which is qualitatively different from the 1D
case. Another interesting phenomenon we found was that, some
defect-mode branches bifurcate \emph{not} from Bloch-band edges, but
from quasi-edge points within Bloch bands. When the defect is
localized but strong, defect modes were studied numerically. It was
found that both the repulsive and attractive defects can guide
various types of defect modes such as fundamental, dipole, tripolar,
quadrupole, and vortex modes. These modes reside in various bandgaps
of the photonic lattice. As the defect strength increases, defect
modes move from lower bandgaps to higher ones when the defect is
repulsive, but remain within each bandgap when the defect is
attractive. The same phenomena are observed when the defect is held
fixed while the applied dc field increases. If the defect is
non-localized (i.e. the defect does not disappear at infinity in the
lattice), we showed that DMs exhibit two new features: (i) DMs can
be embedded inside the continuous spectrum; (ii) DMs can bifurcate
out from edges of the continuous spectrum algebraically rather than
exponentially. These theoretical results pave the way for further
experimental demonstrations of various type of DMs in a 2D lattice.

\section*{Acknowledgments}
We thank Dr. Dmitry Pelinovsky for helpful discussions. This work
was partially supported by the Air Force Office of Scientific
Research and National Science Foundation.


\begin{thebibliography}{200}

\bibitem{Russell} P.St.J. Russell, Science 299, 358 (2003).
\bibitem{Joannopoulos} J.D. Joannopoulos, R.D. Meade, J.N. Winn,  {\it Photonic Crystals:
Molding the Flow of Light} (Princeton University Press, Princeton,
NJ, 1995).
\bibitem{Christodoulides_review} D. N. Christodoulides, F. Lederer, and Y. Silberberg, Nature
(London) 424, 817 (2003)
\bibitem{Kivshar_book} Y. S. Kivshar and G.
P. Agrawal, {\it Optical Solitons: From Fibers to Photonic Crystals}
(Academic Press, New York, 2003).

\bibitem{Segev_nature}
J.W. Fleischer, M. Segev, N.K. Efremidis, and D.N. Christodoulides,
Nature 422, 147 (2003).

\bibitem{ChenPRL04}
H. Martin, E.D. Eugenieva, Z. Chen and D.N. Christodoulides,
``Discrete solitons and soliton-induced dislocations in
partially-coherent photonic lattices," Phys. Rev. Lett. 92, 123902
(2004).

\bibitem{ChenYangPRL} I. Makasyuk, Z. Chen and J. Yang, ``Bandgap guidance in optically-induced photonic lattices with a negative defect",
Phys. Rev. Lett. 96, 223903 (2006).

\bibitem{Chen_OE}
X. Wang, J. Young, Z. Chen, D.Weinstein, and J. Yang, ``Observation
of lower to higher bandgap transition of one-dimensional defect
modes", Opt. Exp. 14, 7362-7367 (2006).


\bibitem{YangOL} F. Fedele, J. Yang, and Z. Chen,
``Defect modes in one-dimensional photonic latices." Opt. Lett. 30,
1506 (2005).

\bibitem{YangChendefect} F. Fedele, J. Yang, and Z. Chen, ``Properties of defect modes in one-dimensional
optically-induced photonic latices." Stud. Appl. Math. 115, 279-301
(2005)

\bibitem{Kivshar_defect}
A. A. Sukhorukov and Y. S. Kivshar, ``Nonlinear localized waves in a
periodic medium", Phys. Rev. Lett. 87, 083901 (2001).

\bibitem{YangChenPRE} J. Yang and Z. Chen,
``Defect solitons in photonic lattices." Phys. Rev. E 73, 026609
(2006).

\bibitem{Silberberg} U. Peschel, R. Morandotti, J. S. Aitchison, H. S. Eisenberg,
and Y. Silberberg, Appl. Phys. Lett. 75, 1348 (1999).

\bibitem{Eisenberg98}
H.S. Eisenberg, Y. Silberberg, R. Morandotti, A. R. Boyd, and J. S.
Aitchison, ``Observation of discrete solitons in optical waveguide
arrays," Phys. Rev. Lett. 81, 3383 (1998).


\bibitem{Neshev_dipole} D.N. Neshev, E. Ostrovskaya, Y. Kivshar, and W. Krolikowski,
``Spatial solitons in optically induced gratings", Opt. Lett. 28,
710 (2003).

\bibitem{Yang_dipole} J. Yang, I. Makasyuk, A. Bezryadina, and
Z. Chen, ``Dipole and quadrupole solitons in optically-induced
two-dimensional photonic lattices: theory and experiment." Stud.
Appl. Math. 113, 389 (2004).

\bibitem{Yang_Muss} J. Yang and Z. Musslimani, ``Fundamental and vortex solitons in a two-dimensional optical lattice."
Opt. Lett. 28, 2094 (2003).

\bibitem{Chen_vortex}
D.N. Neshev, T.J. Alexander, E.A. Ostrovskaya, Y.S. Kivshar, H.
Martin, I. Makasyuk, Z. Chen, "Observation of discrete vortex
solitons in optically-induced photonic lattices," Phys. Rev. Lett.
92, 123903 (2004).

\bibitem{Segev_vortex} J.W. Fleischer, G. Bartal, O. Cohen, O. Manela, M. Segev, J.
Hudock, D.N. Christodoulides, "Observation of vortex-ring discrete
solitons in 2D photonic lattices." Phys. Rev. Lett. 92, 123904
(2004).

\bibitem{Segev_highervortex}
G. Bartal, O. Manela, O. Cohen, J.W. Fleischer, and M. Segev,
``Observation of Second-Band Vortex Solitons in 2D Photonic
Lattices", Phys. Rev. Lett. 95, 053904 (2005).

\bibitem{Kivshar_C_onemode}
R. Fischer, D. Trager, D.N. Neshev, A.A. Sukhorukov, W. Krolikowski,
C. Denz, and Y.S. Kivshar, ``Reduced-Symmetry Two-Dimensional
Solitons in Photonic Lattices", Phys. Rev. Lett. 96, 023905 (2006).

\bibitem{Shi} Z. Shi, J. Yang and Z. Chen, ``Solitary Waves Bifurcated from Bloch Bands in Two-dimensional Periodic
Media", preprint.

\bibitem{Kartashov} Y.V. Kartashov, V.A. Vysloukh, and L. Torner,
``Rotary Solitons in Bessel Optical Lattices", Phys. Rev. Lett. 93,
093904 (2004).

\bibitem{Chen_ring}
X. Wang, Z. Chen, and P. G. Kevrekidis, ``Observation of discrete
solitons and soliton rotation in periodic ring lattices", Phys. Rev.
Lett. 96, 083904 (2006).

\bibitem{Ablowitz} M.J. Ablowitz, B. Ilan, E. Schonbrun, and R.
Piestun, ``Solitons in two-dimensional lattices possessing defects,
dislocations, and quasicrystal structures", Phys. Rev. E 74, 035601
(2006).

\bibitem{Peli_1D} D. E. Pelinovsky, A. A. Sukhorukov, and Y. S. Kivshar,
``Bifurcations and stability of gap solitons in periodic
potentials", Phys. Rev. E 70, 036618 (2004).

\bibitem{Christodoulides_model} D. N. Christodoulides and M. I. Carvalho,
J. Opt. Soc. Am. B 12, 1628 (1995).

\bibitem{Segev_model} M. Segev, M. Shih, and G. C. Valley, J. Opt. Soc. Am. B 13, 706
(1996).

\bibitem{Peli_exp} D.E. Pelinovsky and C. Sulem, ``Asymptotic
approximations for a new eigenvalue in linear problems without a
threshold", Theor. Math. Phys. 122, 98-106 (2000).

\bibitem{YangLakoba} J. Yang and T.I. Lakoba, ``Universally-convergent squared-operator iteration methods for solitary waves in general nonlinear wave
equations", Stud. Appl. Math.  118, 153-197 (2007).

\bibitem{keller} F. Odeh and J.B. Keller,
``Partial differential equations with periodic coefficients and
Bloch waves in crystals.'' J. Math. Phys. 5, 1499 (1964).

\bibitem{Beketov}
F. S. Rofe-Beketov, ``A test for the finiteness of the number of
discrete levels introduced into the gaps of a continuous spectrum by
perturbations of a periodic potential", Soviet Math. Dokl. 5,
689-692 (1964).

\bibitem{Kuchment} P. Kuchment and B. Vainberg, ``On absence of embedded eigenvalues for
Schrödinger operators with perturbed periodic potentials", Commun.
Part. Diff. Equat. 25, 1809 - 1826 (2000).

\end{thebibliography}
\end{document}